\documentclass[journal]{IEEEtran}

\usepackage{array, calc, color, multicol}
\usepackage{algorithm}
\usepackage[noend]{algorithmic} 
\usepackage{graphics, epstopdf, graphicx, epsfig, psfrag, epsf, subfigure}
\usepackage{amssymb, amsfonts, amsmath, times}
\usepackage{multicol,balance, verbatim}
\usepackage{textcomp, mathrsfs, stmaryrd, setspace}

\newcommand{\ie}{{\em i.e., }}
\newcommand{\eg}{{\em e.g., }}
\newcommand{\Eg}{{\em E.g., }}

\newtheorem{example}{Example}

\DeclareGraphicsExtensions{.eps, .pdf, .png, .jpg}   

\newcommand{\Sset}{\mathcal{S}}

\newcommand{\Nset}{\mathcal{N}}

\newcommand{\Lset}{\mathcal{L}}

\DeclareMathOperator*{\argmax}{arg\,max}

\IEEEoverridecommandlockouts

\begin{document}

\title{Separation of Routing and Scheduling in Backpressure-Based Wireless Networks
\thanks{This work was supported by NSF grant CNS-0915988, ONR grant N00014-12-1-0064, ARO Muri grant number W911NF-08-1-0238. The preliminary results of this paper were presented in part at the IEEE Conference on Computer Communications (INFOCOM), Turin, Italy, April 2013.}
\thanks{H.~Seferoglu was with the Laboratory for Information and Decision Systems (LIDS), Massachusetts Institute of Technology. She is currently with the Electrical and Computer Engineering Department at University of Illinois at Chicago. Email: {\tt hulya@uic.edu}. Mail: 1037 Science and Engineering Offices (M/C 154), 851 South Morgan Street, Chicago, IL, 60607.}
\thanks{E.~Modiano is with the Laboratory for Information and Decision Systems (LIDS), Massachusetts Institute of Technology. Email: {\tt modiano@mit.edu}. Mail: 77 Massachusetts Avenue, Room 33-412, Cambridge, MA 02139.}
}

\author{Hulya Seferoglu, {\em Member}, {\em IEEE}, Eytan Modiano, {\em Fellow}, {\em IEEE}}

\maketitle

\begin{abstract}
Backpressure routing and scheduling, with its throughput-optimal operation guarantee, is a promising technique to improve throughput in wireless multi-hop networks. Although backpressure is conceptually viewed as layered, the decisions of routing and scheduling are made jointly, which imposes several challenges in practice. In this work, we present Diff-Max, an approach that separates routing and scheduling and has three strengths: (i) Diff-Max improves throughput significantly, (ii) the separation of routing and scheduling makes practical implementation easier by minimizing cross-layer operations; \ie routing is implemented in the network layer and scheduling is implemented in the link layer, and (iii) the separation of routing and scheduling leads to modularity; \ie routing and scheduling are independent modules in Diff-Max, and one can continue to operate even if the other does not. Our approach is grounded in a network utility maximization (NUM) formulation and its solution. Based on the structure of Diff-Max, we propose two practical schemes: Diff-subMax and wDiff-subMax. We demonstrate the benefits of our schemes through simulation in ns-2. 
\end{abstract}


\section{\label{sec:intro}Introduction}
Backpressure routing and scheduling has emerged from the pioneering work in \cite{tass_eph1}, \cite{tass_eph2}, which showed that, in wireless networks, one can stabilize queues for any feasible traffic by making routing and scheduling decisions based on queue backlog differences. Moreover, it has been shown that backpressure can be combined with flow control to provide utility-optimal operation guarantee \cite{neely_mod}.

The strengths of these techniques have recently increased the interest in practical implementation of backpressure in wireless networks, some of which are summarized in Section \ref{sec:related}. However, the practical implementation of backpressure imposes several challenges mainly due to the joint nature of the routing and scheduling, which is the focus of this paper.

In the backpressure framework, each node constructs per-flow queues. Based on the per-flow queue backlog differences, and by taking into account the state of the network, each node makes routing and scheduling decisions (note that scheduling algorithm is also called as max-weight \cite{neely_book}). Although the backpressure framework is conceptually viewed as layered, the decisions of routing and scheduling are made jointly. To better illustrate this point, let us discuss the following example.

\begin{figure}[t!]
\vspace{0pt}
\centering
\subfigure[Backpressure]{ \label{fig:intro_example_a} \scalebox{.66}{\includegraphics{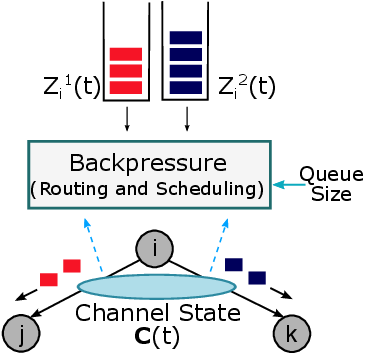}} }
\subfigure[Diff-Max] {\label{fig:intro_example_b} \scalebox{.66}{\includegraphics{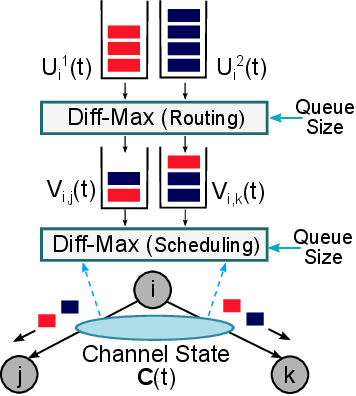}} }
\caption{Example topology consisting of three nodes; $i$, $j$, $k$, and two flows; $1$, $2$. Note that this small topology is a zoomed part of a large multi-hop wireless network. The source and destination nodes of flows $1$ and $2$ are not shown in this example, \ie nodes $i$, $j$, $k$ are intermediate nodes which route and schedule flows $1$ and $2$. (a) Backpressure: Node $i$ constructs per-flow queues; $Z_{i}^{1}$ and $Z_{i}^{2}$, and determines the queue backlog differences at time $t$; $D_{i,j}^{s}(t) = Z_{i}^{s}(t)-Z_{j}^{s}(t)$, $D_{i,k}^{s}(t) = Z_{i}^{s}(t)-Z_{k}^{s}(t)$, where $s \in \{1,2\}$. Based on the differences as well as the channel state of the network, $\boldsymbol C(t)$, backpressure makes joint routing and scheduling decisions. (b) Diff-Max: Node $i$ constructs per-flow queues; $U_{i}^{1}$ and $U_{i}^{2}$ in the network layer, and per-link queue sizes; $V_{i,j}$ and $V_{i,k}$ in the link layer, and makes routing decision based on the queue backlog differences at time $t$; $\tilde{D}_{i,j}^{s}(t) = U_{i}^{s}(t)-U_{j}^{s}(t) - V_{i,j}(t)$, $\tilde{D}_{i,k}^{s}(t) = U_{i}^{s}(t)-U_{k}^{s}(t) - V_{i,k}(t)$, where $s \in \{1,2\}$. Separately, node $i$ makes the scheduling decision based on $V_{i,j}(t)$, $V_{i,k}(t)$, and $\boldsymbol C(t)$. }
\vspace{-10pt}
\label{fig:example_backpressure_diffmax}
\end{figure}

\begin{example}\label{ex1}
Let us consider Fig.~\ref{fig:example_backpressure_diffmax}(a) for backpressure operation. At time $t$, node $i$ makes routing and scheduling decisions for flows $1$ and $2$ based on the per-flow queue sizes; $Z_{i}^{1}(t)$, $Z_{i}^{2}(t)$, as well as the queue sizes of the other nodes, \ie $j$ and $k$ in this example, and using the channel state of the network $\boldsymbol C(t)$. In particular, backpressure determines a packet (and its flow) that should be transmitted over link $i-j$ by $s^{*} = \argmax \{D_{i,j}^{1}(t),D_{i,j}^{2}(t)\}$ such that $s^{*} \in \{1,2\}$. The decision mechanism is the same for link $i-k$. 
The scheduling algorithm also determines the link activation policy. In particular, the maximum backlog difference over each link is calculated as; $D_{i,j}^{*}(t)=D_{i,j}^{s^{*}}(t)$ and $D_{i,k}^{*}(t)=D_{i,k}^{s^{*}}(t)$. Based on $D_{i,j}^{*}(t), D_{i,k}^{*}(t)$ and $\boldsymbol C(t)$, the scheduling algorithm determines the link that should be activated. Note that the decisions of routing and scheduling are made jointly in backpressure, which imposes several challenges in practice. We elaborate on them next.
\hfill $\Box$
\end{example}

Routing algorithms are traditionally designed in the network layer, while the scheduling algorithms are implemented in the link layer. However, the joint routing and scheduling nature of backpressure imposes challenges for practical implementation. To deal with these challenges, \cite{DiffQ} implements backpressure at the link layer, and \cite{umut_stolyar} proposes updates to the MAC layer. This approach is practically difficult due to device memory limitations and strict limitations imposed by device firmware and drivers not to change the link layer functionalities. The second approach is to implement backpressure in the network layer, \eg \cite{horizon}, \cite{routing_wtht_routes}, \cite{xpress}, which requires joint operation of the network and link layers so that backpressure implemented in the network layer operates gracefully with the link layer functionalities. Thus, the network and link layers should work together synchronously, which may not be practical for many off-the-shelf devices.

Existing networks are designed in layers, in which protocols and algorithms are modular and operate independently at each layer of the protocol stack. \Eg routing algorithms at the network layer should work in a harmony with different types of scheduling algorithms in the link layer. However, the joint nature of backpressure stresses joint operation and hurts modularity, which is especially important in contemporary wireless networks, which may vary from a few node networks to ones with hundreds of nodes. It is natural to expect that different types of networks, according to their size as well as software and hardware limitations, may choose to employ backpressure partially or fully. \Eg some networks may be able to employ both routing and scheduling algorithms, while others may only employ routing. Therefore, the algorithms of backpressure, \ie routing and scheduling should be modular.

In this paper, we are interested in a framework in which the routing and scheduling are separated. We seek to find such a scheme where the routing operates in the network layer and the scheduling is implemented in the link layer. The key ingredients of our framework, which we call Diff-Max\footnote{Note that {\em Diff} means that the routing is based on the queue {\em diff}erences, and {\em Max} refers to the fact that the scheduling is based on the {\em max}imum of the (weighted) link layer queues. Finally, the hyphen in Diff-Max is to mention the separation of the routing and scheduling.}, are; (i) per-flow queues at the network layer and making routing decisions based on their differences, (ii) per-link queues at the link layer and making scheduling decisions based on their size.

{\em Example 1 - continued:}
Let us consider Fig.~\ref{fig:example_backpressure_diffmax}(b) for Diff-Max operation. (i) Routing: at time $t$, node $i$ makes routing decision for flows $1$ and $2$ based on queue backlogs $\tilde{D}_{i,j}^{s}(t)$ and $\tilde{D}_{i,k}^{s}(t)$. This decision is made at the network layer and the routed packets are inserted into the link layer queues. Note that in backpressure, routed packets are scheduled jointly, \ie when a packet is routed, it should be transmitted if the corresponding links are activated. Hence, both algorithms should make the decision jointly in backpressure. However, in Diff-Max, a packet may be routed at time $t$, and scheduled and transmitted at a later time $t+T$ where $T>0$. (ii) Scheduling: at the link layer, links are activated and packets are transmitted based on per-link queue sizes; $V_{i,j}$, $V_{i,k}$, and $\boldsymbol C(t)$. The details of Diff-Max are provided in Section~\ref{sec:opt2}.
\hfill $\Box$

Our approach is grounded in a network utility maximization (NUM) framework \cite{tutorial_doyle}. The solution decomposes into several parts with an intuitive interpretation, such as routing, scheduling, and flow control. The structure of the NUM solution provides insight into the design of our scheme, Diff-Max. By separating routing and scheduling, Diff-Max makes the practical implementation easier and minimizes cross-layer operations. 
The following are the key contributions of this work:
\begin{itemize}
  \item We propose a new system model and NUM framework to separate routing and flow scheduling. Our solution to the NUM problem, separates routing and scheduling such that routing is implemented at the network layer, and scheduling is at the link layer. Based on the structure of the NUM solution, we propose Diff-Max. We show that the deterministic version of Diff-Max optimizes utility, and we conjecture that its stochastic version satisfies stability and utility optimality.
  \item We extend Diff-Max to employ routing and intra-node scheduling, but disable inter-node scheduling. We call the new framework Diff-subMax, which reduces computational complexity and overhead significantly, and provides high throughput improvements in practice. Namely, Diff-subMax only needs information from one-hop away neighbors to make its routing and scheduling decisions. Furthermore, we show that the deterministic version of Diff-subMax provides utility optimality for the networks with  pre-determined inter-node scheduling.
  \item We propose a window-based routing scheme, wDiff-subMax, which implements routing, but disables the scheduling. wDiff-subMax is a heuristic developed based on Diff-Max and Diff-subMax, and it is designed for the scenarios, in which the implementation of the scheduling in the link layer is impossible (or not desirable) \eg due to device restrictions. wDiff-subMax makes the routing decisions on the fly, and reduces overhead.
  \item We evaluate our schemes in a multi-hop setting and consider their interaction with transport, network, and link layers. 
      We implement our schemes in a simulator; ns-2 \cite{ns2}, and show that they significantly improve throughput as compared to adaptive routing schemes such as Ad hoc On-Demand Distance Vector (AODV) \cite{aodv} and Destination-Sequenced Distance-Vector Routing (DSDV) \cite{dsdv}. 
\end{itemize}

The structure of the rest of the paper is as follows. Section~\ref{sec:system} gives an overview of the system model. Section~\ref{sec:opt2} presents the Diff-Max formulation and design.
Section~\ref{sec:algs} presents the development and implementation details of Diff-Max schemes. 
Section~\ref{sec:performance} presents simulation results. Section~\ref{sec:related} presents related work. Section~\ref{sec:conclusion} concludes the paper.

\section{\label{sec:system}System Overview}
In this section, we provide an overview of the system model for separation of routing and scheduling. We also provide background on the backpressure framework so that we can make a connection and comparison between our scheme and backpressure throughout the rest of the paper.

\subsection{Separation of Routing and Scheduling}
\begin{figure}
\centering
\scalebox{.8}{\includegraphics[bb=0 0 215 205]{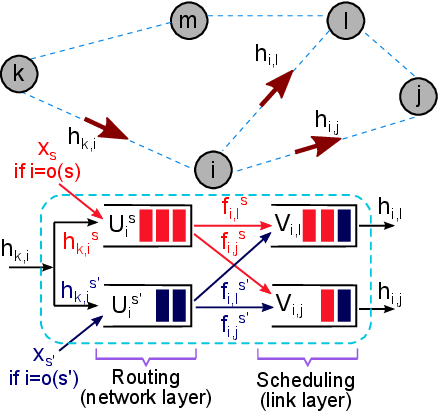}}
\caption{A wireless mesh network. The queues at the network and link layers as well as the interaction among the queues inside node $i$ are shown in detail. $U_{i}^{s}$ and $U_{i}^{s'}$ are the network layer queues for flows $s$ and $s'$, and $V_{i,j}$ and $V_{i,l}$ are the per-link queues for the links; $i-j$, $i-l$. The routing algorithm operates in the network layer, the scheduling is implemented in the link layer. }
\label{fig:main-example}
\vspace{-20pt}
\end{figure}

We consider multi-hop wireless networks, in which packets from a source traverse potentially multiple wireless hops before being received at their destination. In this setup, each wireless node is able to perform routing, scheduling, and flow control. In this section, we provide an overview of this setup and highlight some of its key characteristics. Fig.~\ref{fig:main-example} shows the key parts of our system model in an example topology.

{\em Setup:} We consider a wireless network which consists of $N$ nodes and $L$ edges, where $\Nset$ is the set of nodes and $\Lset$ is the set of edges. We consider in our formulation and analysis that time is slotted, and $t$ refers to the beginning of slot $t$.

{\em Sources and Flows:} Let $\Sset$ be the set of unicast flows between source-destination pairs in the network. Each flow $s \in \Sset$ arrives from an application layer to a transport layer with rate $A_s(t)$, $\forall s \in \Sset$ at time slot $t$. The arrivals are i.i.d. over the slots and their expected values are; $\lambda_s = E[A_s(t)]$, and $E[A_s(t)^{2}]$ are finite. The transport layer stores the arriving packets in reservoirs (\ie transport layer per-flow queues), and controls the flow. In particular, each source $s$ is associated with rate $x_{s}$ and a utility function $g_{s}(x_{s})$, which we assume to be a strictly concave function of $x_{s}$. The transport layer determines $x_{s}(t)$ at time slot $t$ according to the utility function $g_{s}(x)$, and $x_{s}(t)$ packets are transmitted from the transport layer reservoir to the network layer at slot $t$. 

{\em Queue Structures:} At node $i \in \Nset$, there are network and link layer queues. The network layer queues are per-flow queues; \ie $U_{i}^{s}$ is the queue at node $i \in \Nset$ that only stores the packets from flow $s \in \Sset$. The link layer queues are per-link queues; \ie at each node $i \in \Nset$, a link layer queue $V_{i,j}$ is constructed for a neighbor node $j \in \Nset$ (Fig.~\ref{fig:main-example}).\footnote{Note that in some devices, there might be only one queue (per-node queue) for data transmission instead of per-link queues in the link layer. Developing a model with per-node queues is challenging due to coupling among actions and states, so it is an open problem.}

{\em Flow Rates:} Our model optimizes the flow rates among different nodes as well as the flow rates in a node among different layers; transport, network, and link layer.

The transport layer determines $x_{s}(t)$ at time $t$, and passes $x_{s}(t)$ packets to the network layer. These packets are inserted in the network layer queue; $U_{i}^{s}$ (assuming that node $i$ is the source node of flow $s$).

The flow rate from the network layer to the link layer queues is $f_{i,j}^{s}(t)$. In particular, $f_{i,j}^{s}(t)$ is the flow rate of the packets, belonging to flow $s$, from the network layer queue; $U_{i}^{s}$ to the link layer queue; $V_{i,j}$ at node $i$. Note that the optimization of flow rate $f_{i,j}^{s}(t)$ is the routing decision, since it basically determines how many packets from flow $s$ should be forwarded/routed to node $j$.

The link transmission rate from $i$ to $j$ is $h_{i,j}(t)$. Note that $h_{i,j}(t)$ bounds per-flow data rates; \ie $h_{i,j}(t) \geq \sum_{s \in \Sset} h_{i,j}^{s}(t)$. \Eg $h_{k,i}(t) \geq h_{k,i}^{s}(t) + h_{k,i}^{s'}(t)$ in Fig.~\ref{fig:main-example} where $h_{k,i}^{s}(t)$ is the flow rate of flow $s$ over link $k-i$. Note that the optimization of link transmission rate $h_{i,j}(t)$ corresponds to the scheduling decisions, since it determines which packets from which link layer queues should be transmitted as well as whether a link is activated.

At every time slot $t$, $U_{i}^{s}$ changes according to the following dynamics.
\begin{align} \label{eq:queue_U}
& U_{i}^{s}(t+1) \leq \max [U_{i}^{s}(t) - \sum_{j \in \Nset} f_{i,j}^{s}(t), 0] + \sum_{j \in \Nset} h_{j,i}^{s}(t) \nonumber \\
& + x_{s}(t)1_{[i=o(s)]}
\end{align} where $o(s)$ is the source node of flow $s$ and $1_{[i=o(s)]}$ is an indicator function, which is $1$ if $i=o(s)$, and $0$, otherwise. Note that (\ref{eq:queue_U}) is an inequality, because the actual amount of flow rate of flow $s$ over link $j-i$ may be lower than $h_{j,i}^{s}(t)$ as there may not be enough packets from flow $s$ in the link layer queue at node $j$.

At every time slot $t$, $V_{i,j}$ changes according to the following queue dynamics.
\begin{align} \label{eq:queue_V}
V_{i,j}(t+1) \leq \max [V_{i,j}(t) - h_{i,j}(t), 0] + \sum_{s \in \Sset} f_{i,j}^{s}(t)
\end{align} Note that (\ref{eq:queue_V}) is an inequality as the number of packets in $U_{i}^{s}(t)$ may be lower than $f_{i,j}^{s}(t)$.

{\em Channel Model:} At slot $t$, $\boldsymbol C(t)$ is the channel state vector, where $\boldsymbol C(t) = \{C_{1}(t), ..., C_{l}(t), ..., C_{L}(t)\}$, where $l$ represents the edges such that $l = (i,j)$, $(i,j) \in \Lset$ and $i \neq j$. We assume that $C_{l}(t)$ is the state of link $l$ at time $t$ and takes values from the set $\{ON,OFF\}$ according to a probability distribution which is i.i.d. over time slots. If  $C_{l}(t) = ON$, packets can be transmitted with rate $R_l$. Otherwise; (\ie if $C_{l}(t) = OFF$), packets cannot be transmitted successfully.

Let $\Gamma_{\boldsymbol C(t)}$ denote the set of the link transmission rates feasible at time slot $t$ for channel state $\boldsymbol C(t)$ accounting for interference among the wireless links. In particular, at every time slot $t$, the link transmission vector $\boldsymbol h(t) = \{h_1(t), ..., h_l(t), ... h_L(t)\}$ should be constrained such that $\boldsymbol h(t)$ $\in \Gamma_{\boldsymbol C(t)}$. 

{\em Stability Region:}
Let $(\lambda_s)$ be the vector of arrival rates $\forall s \in \Sset$. The network stablity region $\Lambda$ is defined as the closure of all arrival rate vectors that can be stably transmitted in the network, considering all possible routing and scheduling policies \cite{tass_eph1}, \cite{tass_eph2}, \cite{neely_mod}. $\Lambda$ is fixed and depends only on channel statistics and interference.

\subsection{Background on Backpressure}
\begin{figure}
\centering
\scalebox{.8}{\includegraphics{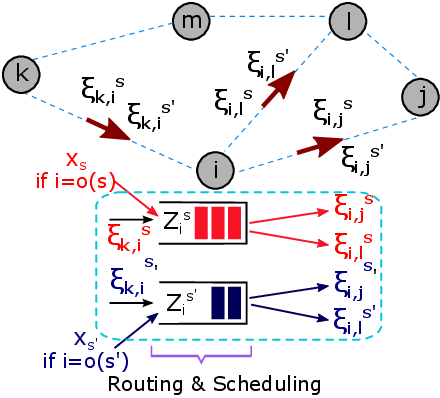}}
\caption{Backpressure system model in a wireless mesh network. Per-flow queues inside node $i$ are shown in detail. $Z_{i}^{s}$ and $Z_{i}^{s'}$ are the per-flow queues for flows $s$ and $s'$, respectively. The backpressure routing \& scheduling algorithm operates jointly over the per-flow queues. }
\label{fig:main-example-backpressure}
\vspace{-20pt}
\end{figure}

In this section, we provide background on backpressure so that we can make a connection and comparison between our scheme and backpressure throughout the rest of the paper. We consider a similar system model as in the previous section. 
Fig.~\ref{fig:main-example-backpressure} shows the key parts of backpressure system model in an example topology.

At node $i \in \Nset$, there are per-flow queues. The per-flow queue $Z_{i}^{s}$ is the queue at node $i \in \Nset$ that only stores the packets from flow $s \in \Sset$. The flow rate $x_{s}(t)$ is determined at time $t$, and the corresponding number of packets are inserted in the network layer queue; $Z_{i}^{s}$ (assuming that node $i$ is the source node of flow $s$). The flow rate from node $i$ to node $j$ for flow $s$ is $\xi_{i,j}^{s}$. 

Note that the optimization of flow rate $\xi_{i,j}^{s}(t)$ is both the routing and scheduling decision, since it basically determines how many packets from flow $s$ should be forwarded/routed over which links. Thus, at every time slot $t$, $Z_{i}^{s}$ changes according to the following dynamics.
\begin{align} \label{eq:queue_U_tilde}
& Z_{i}^{s}(t+1) \leq \max [Z_{i}^{s}(t) - \sum_{j \in \Nset} \xi_{i,j}^{s}(t), 0] + \sum_{j \in \Nset} \xi_{j,i}^{s}(t) \nonumber \\
& + x_{s}(t)1_{[i=o(s)]}
\end{align}  The backpressure scheme operates on per-flow queues $Z_{i}^{s}$ and makes routing and scheduling decisions based on the following algorithm. 

\underline{Backpressure:}
\begin{itemize}
 \item {\em Routing \& Scheduling:} At each time slot $t$, the rate $\xi_{i,j}^{s}(t)$ is determined by;
 \begin{align} \label{eq_stoc:routing_scheduling_backpressure}
  \max_{\boldsymbol \xi} &  \sum_{j \in \Nset_{i}} \sum_{s \in \Sset} \xi_{i,j}^{s}(t) (Z_{i}^{s}(t) - Z_{j}^{s}(t)) \nonumber \\
  \mbox{s.t. } &  \boldsymbol \xi(t) \in \Gamma_{\boldsymbol C(t)},
  \end{align}
  
\end{itemize} Backpressure routing and scheduling algorithm in (\ref{eq_stoc:routing_scheduling_backpressure}) stabilizes the network and average queue backlog sizes are bounded \cite{tass_eph1}, \cite{tass_eph2}. Moreover, it has been shown that backpressure can be combined with flow control to provide utility-optimal operation guarantee \cite{neely_mod}.

\section{\label{sec:opt2}Diff-Max: Formulation and Design}
\subsection{Network Utility Maximization}
In this section, we formulate and design Diff-Max. Our first step is the NUM formulation of the problem and its solution. This approach sheds light into the structure of the Diff-Max algorithms. \footnote{NUM optimizes the average values of the parameters (\ie flow rates) that are defined in Section~\ref{sec:system}. By abuse of notation, we use a variable, \eg $\phi$ as the average value $\phi(t)$ in our NUM formulation if both $\phi$ and $\phi(t)$ refers to the same parameter.}

\subsubsection{Formulation}
Our objective is to maximize the total utility by optimally choosing the flow rates $x_s$, as well as the amount of data traffic that should be routed to each neighbor node; \ie $f_{i,j}^{s}$, and the link transmission rates; \ie $h_{i,j}$.
\begin{align} \label{opt:eq1}
\max_{\boldsymbol x, \boldsymbol f, \boldsymbol h} & \sum_{s \in \Sset} g_{s}(x_{s}) \nonumber \\
\mbox{s.t. } & \sum_{j \in \Nset} f_{i,j}^{s} - \sum_{j \in \Nset} h_{j,i}^{s} = x_{s}1_{[i=o(s)]}, \forall i \in \Nset, s \in \Sset \nonumber \\
& \sum_{s \in \Sset} f_{i,j}^{s} \leq h_{i,j}, \forall (i,j) \in \Lset \nonumber \\
& f_{i,j}^{s} = h_{i,j}^{s}, \forall s \in \Sset, (i,j) \in \Lset \nonumber \\
& \boldsymbol h \in \tilde{\Gamma}.
\end{align}
The first constraint is the flow conservation constraint at the network layer: at every node $i$ and for each flow $s$, the sum of the total incoming traffic, \ie $\sum_{j \in \Nset} h_{j,i}^{s}$ and exogenous traffic, \ie $x_s$ should be equal to the total outgoing traffic from the network layer, \ie $\sum_{j \in \Nset} f_{i,j}^{s}$. The second constraint is the flow conservation constraint at the link layer; the link transmission rate; \ie $h_{i,j}$ should be larger than the incoming traffic; \ie $\sum_{s \in \Sset} f_{i,j}^{s}$. Note that this constraint is an inequality, because the link transmission rate can be larger than the actual data traffic. The third constraint gives the relationship between the network and link layer per-flow data rates, and the last constraint requires that the vector of link transmission rates, $\boldsymbol h = \{h_1, ..., h_l, ... h_L\}$ should be the element of the available link rates; $\tilde{\Gamma}$. Note that $\tilde{\Gamma}$ is different than $\Gamma_{\boldsymbol C(t)}$ in the sense that $\tilde{\Gamma}$ represents long-term average rates rather than instantaneous rates.

The first two constraints are key to our work, because they determine the incoming and outgoing flow relationships at the network and link layers, respectively. This approach separates routing and scheduling, and assigns the routing to the network layer and scheduling to the link layer. Note that if these constraints are combined in such a way that incoming rate from a node and exogenous traffic should be smaller than the outgoing traffic for each flow, we obtain the backpressure solution \cite{lin_schroff_tutorial}, \cite{lin_schroff_paper}.

\subsubsection{Solution}
Lagrangian relaxation of the first constraint gives the following Lagrange function:
\begin{align} \label{opt:eq1_Lagrange1}
& L(\boldsymbol x, \boldsymbol f, \boldsymbol h, \boldsymbol u, \boldsymbol v) = \sum_{s \in \Sset} g_{s}(x_{s}) + \sum_{i \in \Nset} \sum_{s \in \Sset} u_{i}^{s} \Bigl(\sum_{j \in \Nset} f_{i,j}^{s} -  \nonumber \\
& \sum_{j \in \Nset} h_{j,i}^{s} - x_{s}1_{[i=o(s)]} \Bigr) - \sum_{(i,j) \in \Lset} v_{i,j} \Bigl( \sum_{s \in \Sset} f_{i,j}^{s} - h_{i,j} \Bigr),
\end{align} where $u_{i}^{s}$ and $v_{i,j}$ are the Lagrange multipliers. The Lagrange function can be re-written as;

\begin{align} \label{opt:eq1_Lagrange2}
& L(\boldsymbol x, \boldsymbol f, \boldsymbol h, \boldsymbol u, \boldsymbol v) = \sum_{s \in \Sset} (g_{s}(x_{s}) - u_{o(s)}^{s}x_{s} ) + \sum_{i \in \Nset} \sum_{s \in \Sset} \sum_{j \in \Nset} u_{i}^{s} f_{i,j}^{s} \nonumber \\
& - \sum_{i \in \Nset} \sum_{s \in \Sset} \sum_{j \in \Nset} u_{j}^{s}h_{i,j}^{s} - \sum_{(i,j) \in \Lset} \sum_{s \in \Sset} v_{i,j} f_{i,j}^{s} + \sum_{(i,j) \in \Lset} v_{i,j} h_{i,j}
\end{align} (\ref{opt:eq1_Lagrange2}) can be decomposed into several intuitive sub-problems such as flow control, routing, and scheduling.
First, we solve the Lagrangian with respect to $x_s$:
\begin{equation} \label{opt:eq1_rateControl}
\textstyle x_s = ({g'_{s}})^{-1} \left( u_{o(s)}^{s} \right),
\end{equation} where $({g'_{s}})^{-1}$ is the inverse function of the derivative of $g_{s}$. This part of the solution can be interpreted as the flow control.
Second, we solve the Lagrangian for $f_{i,j}^{s}$ and $h_{i,j}^{s}$. The following part of the solution can be interpreted as the routing.
\begin{align} \label{opt:eq1_routing1}
\max_{\boldsymbol f} & \sum_{i \in \Nset} \sum_{s \in \Sset} \sum_{j \in \Nset} (u_{i}^{s} f_{i,j}^{s} -  u_{j}^{s}h_{i,j}^{s}) - \sum_{(i,j) \in \Lset} \sum_{s \in \Sset} v_{i,j} f_{i,j}^{s} \nonumber \\
\mbox{s.t. } & f_{i,j}^{s} = h_{i,j}^{s}, \mbox{   } \forall i \in \Nset, j \in \Nset, s \in \Sset
\end{align} The above problem is equivalent to;
\begin{align} \label{opt:eq1_routing2}
\max_{\boldsymbol f} & \sum_{(i,j) \in \Lset} \sum_{s \in \Sset} f_{i,j}^{s} (u_{i}^{s} - u_{j}^{s} - v_{i,j})
\end{align}
Third, we solve the Lagrangian for $h_{i,j}$. The following part of the solution can be interpreted as the scheduling.
\begin{align} \label{opt:eq1_scheduling1}
\max_{\boldsymbol h} & \sum_{(i,j) \in \Lset} v_{i,j} h_{i,j} \nonumber \\
\mbox{s.t. } & \boldsymbol h \in \tilde{\Gamma}. 
\end{align}

The decomposed parts of the Lagrangian, \ie Eqs.~(\ref{opt:eq1_rateControl}), (\ref{opt:eq1_routing2}), (\ref{opt:eq1_scheduling1}) and the Lagrange multipliers; $u_{i}^{s}$ and $v_{i,j}$ can be solved iteratively via a gradient descent algorithm. The convergence properties of this solution to the utility optimal operating point are provided in Appendix A. 
Next, we design Diff-Max based on the structure of the NUM solution.

\subsection{Diff-Max}
Now, we provide a stochastic control strategy including routing, scheduling, and flow control. The strategy, \ie Diff-Max, which mimics the NUM solution, combines separated routing and scheduling together with the flow control.

\underline{Diff-Max:}
\begin{itemize} 
 \item {\em Routing:} Node $i$ determines $f_{i,j}^{s}(t)$ according to;
 \begin{align} \label{eq_stoc:routing}
  \max_{\boldsymbol f} &  \sum_{j \in \Nset_{i}} \sum_{s \in \Sset} f_{i,j}^{s}(t) (\tilde{U}_{i}^{s}(t) - \tilde{U}_{j}^{s}(t) - V_{i,j}(t)) \nonumber \\
  \mbox{s.t. } &  \sum_{j \in \Nset_{i}} \sum_{s \in \Sset} f_{i,j}^{s}(t) \leq  F_{i}^{max}
  \end{align} where $F_{i}^{max}$ is constant larger than the maximum outgoing rate from node $i$, $\Nset_{i}$ is the set of node $i$'s neighbors, and $\tilde{U}_{i}^{s}(t)$ is the network layer virtual queue. 
  
According to (\ref{eq_stoc:routing}), $f_{i,j}^{s}(t)$ packets are removed from $\tilde{U}_{i}^{s}(t)$ and inserted to the link layer queue $V_{i,j}(t)$. This routing algorithm mimics (\ref{opt:eq1_routing2}) and has the following interpretation. Packets from flow $s$ can be transmitted to the next hop node $j$ as long as the network layer queue in the next hop (node $j$) is small, which means that node $j$ is able to route the packets, and the link layer queue at the current node (node $i$) is small, which means that the congestion over link $i-j$ is relatively small. If the number of packets in $U_{i}^{s}(t)$ is smaller than the routing variable calculated by (\ref{eq_stoc:routing}), the packets are transmitted to the link layer queues beginning from the largest $U_{i}^{s}(t) - U_{j}^{s}(t) - V_{i,j}(t)$.

  The routing algorithm in (\ref{eq_stoc:routing}) uses per-link queues as well as per-flow queues, which is the main difference of (\ref{eq_stoc:routing}) as compared to the backpressure routing. The backpressure routing only uses per-flow queues, and does not take into account the state of the link layer queues, which do not exist in the standard backpressure formulation. 
  \item {\em Scheduling:} At each time slot $t$, the link rate $h_{i,j}(t)$ is determined by;
  \begin{align} \label{eq_stoc:scheduling}
  \max_{\boldsymbol h} &  \sum_{(i,j) \in \Lset} V_{i,j}(t)h_{i,j}(t) \nonumber \\
  \mbox{s.t. } &  \boldsymbol h(t) \in \Gamma_{\boldsymbol C(t)}, \forall (i,j) \in \Lset
  \end{align} (\ref{eq_stoc:scheduling}) mimics (\ref{opt:eq1_scheduling1}) and has the following interpretation. The link $i-j$ with the largest queue backlog $V_{i,j}$, taking into account the channel state vector, should be activated, and a packet(s) from the corresponding queue, \ie $V_{i,j}$, should be transmitted. Note that the scheduling in (\ref{eq_stoc:scheduling}) is known to be a difficult problem \cite{tutorial_doyle}, \cite{lin_schroff_tutorial}. Therefore, in Section~\ref{sec:algs}, we propose sub-optimal, low-complexity scheduling algorithms that interact well with the routing algorithm in (\ref{eq_stoc:routing}).

  The scheduling algorithm in (\ref{eq_stoc:scheduling}) differs from backpressure in the sense that it is completely independent from the routing. In particular, (\ref{eq_stoc:scheduling}) makes the scheduling decision based on the per-link queues; $V_{i,j}$ and the channel state; $\boldsymbol C(t)$, while backpressure uses maximum queue backlog differences dictated by the routing algorithm. As it is seen, the routing and scheduling are operating jointly in backpressure, while in Diff-Max, these algorithms are separated.
  \item {\em Flow Control:} At every time slot $t$, the flow/rate controller at the transport layer of node $i$ determines the number of packets that should be passed from the transport layer to the network layer according to:
  \begin{align} \label{eq_stoc:rate_control}
  \max_{\boldsymbol x} & \sum_{[s \in \Sset | i=o(s)]} [Mg_{s}(x_s(t)) -  U_{i}^{s}(t) x_{s}(t) ] \nonumber \\
  \mbox{s.t. } &  \sum_{[s \in \Sset  | i=o(s)]} x_{s}(t) \leq R_{i}^{max}
  \end{align} where $R_{i}^{max}$ is a constant larger than the maximum outgoing rate from node $i$, and $M$ is a finite constant, $M>0$. The flow control in our solution mimics (\ref{opt:eq1_rateControl}) as well as the flow control algorithm proposed in \cite{neely_mod}.
\end{itemize}

\section{\label{sec:algs}Implementation Details}
\begin{figure}
\centering
\scalebox{.55}{\includegraphics[bb=0 0 431 245]{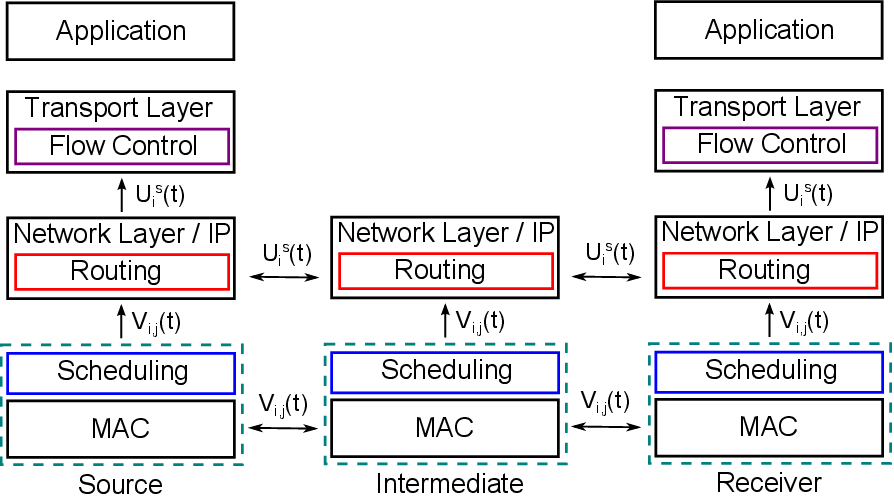}}
\vspace{-5pt}
\caption{Diff-Max operations at end-points and intermediate nodes.}
\label{fig:protocol_stack}
\vspace{-10pt}
\end{figure}

We propose practical implementations of Diff-Max (Fig.~\ref{fig:protocol_stack}) as well as Diff-subMax, which combines the routing algorithm with a sub-optimal scheduling, and wDiff-subMax which makes routing decision based on a window-based algorithm.

\subsection{Diff-Max}
\subsubsection{Flow Control}
The flow control algorithm, implemented at the transport layer at the end nodes (see Fig.~\ref{fig:protocol_stack}), determines the rate of each flow. We implement our flow control algorithm as an extension of UDP in the ns-2 simulator. 

The flow control algorithm, at the source node $i$, divides time into epochs (virtual slots) such as $t_{i}^{1}, t_{i}^{2}, ..., t_{i}^{k}, ...$, where $t_{i}^{k}$ is the beginning of the $k$th epoch. Let us assume that $t_{i}^{k+1} = t_{i}^{k} + T_{i}$ where $T_{i}$ is the epoch duration.

At time $t_{i}^{k}$, the flow control algorithm determines the rate according to (\ref{eq_stoc:rate_control}). We consider $g_{s}(x_s(t)) = \log(x_s(t))$ (note that any other concave utility function can be used).
After $x_{s}(t_{i}^{k})$ is determined, a corresponding number of packets are passed to the network layer, and inserted to the network layer queue $U_{i}^{s}$. 
Packets that are not forwarded to the network layer are stored in a reservoir at the transport layer, and transmitted in later slots. At the receiver node, the transport protocol receives packets from the lower layers and passes them to the application.

\subsubsection{Routing}
The routing algorithm, implemented at the network layer of each node (see Fig.~\ref{fig:protocol_stack}), determines routing policy, \ie the next hop(s) that packets are forwarded to.

The first part of our routing algorithm is the neighbor discovery and queue size information exchange. Each node $i$ transmits a message containing the size of its network layer queues; $U_{i}^{s}$. These messages are in general piggy-backed to data packets. The nodes in the network operates in the promiscuous mode. Therefore, each node, let us say node $j$, overhears a packet from node $i$ even if node $i$ transmits the packet to another node, let us say node $k$. Node $j$ reads the queue size information from the data packet that it receives or overhears (thanks to operating on the promiscuous mode). The queue size information is recorded for future routing decisions. Note that when a node hears from another node through direct or promiscuous mode, it classifies it as its neighbor. The neighbor nodes of node $i$ forms a set $\Nset_{i}$. As we mentioned, queue size information is piggy-backed to data packets. However, if there is no data packet for transmission, the node creates a packet to carry queue size information and broadcasts it.

The second part of our routing algorithm is the actual routing decision. Similar to the flow control algorithm, the routing algorithm divides time into epochs; such as $t_{i}^{'1}, t_{i}^{'2}, ..., t_{i}^{'k}, ...$, where $t_{i}^{'k}$ is the beginning of the $k$th epoch at node $i$. Let us assume that $t_{i}^{'k+1} = t_{i}^{'k} + T'_{i}$ where $T'_{i}$ is the epoch duration. Note that we use $t_{i}^{'k}$ and $T'_{i}$ instead of $t_{i}^{k}$ and $T_{i}$, because these two time epochs do not need to be the same nor synchronized.

At time $t_{i}^{'k}$, the routing algorithm checks $U_{i}^{s}(t_{i}^{'k}) - U_{j}^{s}(t_{i}^{'k}) - V_{i,j}(t_{i}^{'k})$ for each flow $s$. Note that $U_{j}^{s}(t_{i}^{'k})$ is not the instantaneous value of $U_{j}^{s}$ at time $t_{i}^{'k}$, but the latest value of $U_{j}^{s}$ heard by node $i$ before $t_{i}^{'k}$. Note also that $V_{i,j}(t_{i}^{'k})$ is the per-link queue at node $i$, and this information should be passed to the network layer for routing decision. According to (\ref{eq_stoc:routing}), $f_{i,j}^{s}(t_{i}^{'k})$ is determined $\forall j \in \Nset_{i}, \forall s \in \Sset$, and $f_{i,j}^{s}(t_{i}^{'k})$ packets are removed from $U_{i}^{s}$ and inserted to the link layer queue $V_{i,j}$ at node $i$. Note that the link layer transmits packets from $V_{i,j}$ only to node $j$, hence the routing decision is completed. The routing algorithm is summarized in Algorithm~\ref{alg:routing_decision}. Note that Algorithm~\ref{alg:routing_decision} considers that there are enough packets in $U_{i}^{s}$ for transmission. If not, the algorithm lists all the links $j \in \Nset_{i}$ in decreasing order, according to the weight; $U_{i}^{s}(t_{i}^{'k}) - U_{j}^{s}(t_{i}^{'k}) - V_{i,j}(t_{i}^{'k})$. Then, it begins to route packets beginning from the link that has the largest weight.

\begin{algorithm}[t!]
 \caption{The routing algorithm at node $i$ at slot $t_{i}^{'k}$. \label{alg:routing_decision}}
\begin{algorithmic}[1]
\begin{footnotesize}
\FOR {$\forall j \in \Nset_{i}, \forall s \in \Sset$}
\STATE Read the network layer queue size information of neighbors: $U_{j}^{s}(t_{i}^{'k})$
\STATE Read the link layer queue size information: $V_{i,j}(t_{i}^{'k})$
\STATE $f_{i,j}^{s}(t_{i}^{'k}) = 0$
\ENDFOR
\STATE $\{j^{*}, s^{*}\} = \argmax_{[j \in \Nset_{i}, s \in \Sset]} \{U_{i}^{s}(t) - U_{j}^{s}(t) - V_{i,j}(t)\}$
\STATE $f_{i,j^{*}}^{s^{*}}(t_{i}^{'k}) = F_{i}^{max}$
\STATE Remove $f_{i,j^{*}}^{s^{*}}(t_{i}^{'k})$ packets from $U_{i}^{s^{*}}$
\STATE Pass $f_{i,j^{*}}^{s^{*}}(t_{i}^{'k})$ packets to the link layer and insert them in $V_{i,j^{*}}$
\end{footnotesize}
\end{algorithmic}
\end{algorithm}

\subsubsection{Scheduling} The scheduling algorithm in (\ref{eq_stoc:scheduling}) assumes that time is slotted. 
Although there are time-slotted system implementations, and also recent work on backpressure implementation over time-slotted wireless networks \cite{xpress}, IEEE 802.11 MAC, an asynchronous medium access protocol without time slots, is the most widely used MAC protocol in current wireless networks. Therefore, we implement our scheduling algorithm (formulated in (\ref{eq_stoc:scheduling})) on top of 802.11 MAC (see Fig.~\ref{fig:protocol_stack}) with the following updates.

The scheduling algorithm constructs per-link queues at the link layer. Node $i$ knows its own link layer queues, $V_{i,j}$, and estimates the loss probability and link rates. Let us consider that $\bar{p}_l$ and $\bar{R}_l$ are the estimated values of $p_l$ and $R_l$, respectively. $\bar{p}_l$ is calculated as one minus the ratio of correctly transmitted packets over all transmitted packets in a time window over link $l$.\footnote{Note that we do not use instantaneous channel states $C_l(t)$ in our implementation, since it is not practical to get this information. Even if one can estimate $C_{l}(t)$ using physical layer learning techniques, $C_{l}(t)$ should be estimated $\forall l \in \Lset$, which is not practical in current wireless networks.} $\bar{R}_l$ is calculated as the average of the recent (in a window of time) link rates over link $l$. $V_{i,j}$, $\bar{p}_{i,j}$, and $\bar{R}_{i,j}$ are piggy-backed to the data packets and exchanged among nodes. Note that this information should be exchanged among all nodes in the network since each node is required to make its own decision based on global information. Also, each node knows the general topology and interfering links.

The scheduling algorithm that we implemented mimics (\ref{eq_stoc:scheduling}). Each node $i$ knows per-link queues, \ie $V_l$, estimated loss probabilities, \ie $\bar{p}_l$, and link rates, \ie $\bar{R}_{l}$, for $l \in \Lset$ as well all maximal independent sets, which consist of links that are not interfering. Let us assume that there are $Q$ maximal independent sets. For the $q$th maximal independent set such that $q=1,...,Q$, the policy vector is; $\boldsymbol \pi_q = \{\pi_q^1, ..., \pi_q^l, ..., \pi_q^L \}$, where $\pi_q^l=1$ if link $l$ is in the $q$th maximal set, and $\pi_q^l=0$, otherwise. Our scheduling algorithm selects $q^{*}$th maximal independent set such that $q^{*} = \argmax_{[\forall q]} \{\sum_{l \in \Lset} V_l(1-\bar{p}_l)\bar{R}_l\pi_q^{l}\}$. Node $i$ calculates $q^{*}$ whenever one of the parameters; $V_l$, $\bar{p}_l$, $\bar{R}_l$ change. If, according to $q^{*}$, node $i$ decides that it should activate one of its links, then it reduces the contention window size of 802.11 MAC so that node $i$ can access the medium quickly and transmit a packet. If node $i$ should not transmit, then the scheduling algorithm tells 802.11 MAC that there are no packets in the queues available for transmission.
Note that in order to complement Diff-Max scheduling, the 802.11 protocol has to be slightly modified.
The scheduling algorithm is summarized in Algorithm~\ref{alg:scheduling}.

\begin{algorithm}[t!]
 \caption{Diff-Max scheduling algorithm at node $i$. \label{alg:scheduling}}
\begin{algorithmic}[1]
\begin{footnotesize}
\IF {$V_l$, $\bar{p}_l$, or $\bar{R}_l$ is updated such that $l \in \Lset$}
\STATE Determine $q^{*}$ such that $q^{*} = \argmax_{[\forall q]} \{\sum_{l \in \Lset} V_l (1-\bar{p}_l)\bar{R}_l\pi_q^{l}\}$
\IF {$\exists (i,j)$ such that $\pi_{q^{*}}^{(i,j)} = 1$, $\forall j \in \Nset_{i}$ }
\STATE Reduce 802.11 MAC contention window size and access the medium
\STATE Transmit a packet from $V_{i,j}$ according to FIFO rule
\ELSE
\STATE Tell 802.11 MAC that there are no packets in the queues available for transmission
\ENDIF
\ENDIF
\end{footnotesize}
\end{algorithmic}
\end{algorithm}

Note that Algorithm~\ref{alg:scheduling} is a hard problem, because it is reduced to  maximum independent set problem, \cite{tutorial_doyle}, \cite{lin_schroff_tutorial}. Furthermore, it introduces significant amount of overhead; each node needs to know every other node's queue sizes and link loss rates. Due to the complexity of the problem and overhead, we implement this algorithm for small topologies over ns-2 for the purpose of comparing its performance with sub-optimal scheduling algorithms, which we describe next.

\subsection{\label{sec:algs_diffsubMax} Diff-subMax}
Diff-subMax is a low complexity and low overhead counterpart of Diff-Max. The flow control and the routing parts of Diff-subMax is exactly the same as in Diff-Max. The only different part is the scheduling algorithm, which uses 802.11 MAC protocol without any changes.
When a transmission opportunity arises according to the underlying 802.11 MAC at time $t$, then the scheduling algorithm of node $i$ calculates weights for all outgoing links to its neighbors. Let us consider link $i-j$ at time $t$. The weight is $\omega_{i,j}(t)=V_{i,j}(t)(1-\bar{p}_{i,j})\bar{R}_{i,j}$. Based on the weights, the link is chosen as; $l^* = \argmax_{[j \in \Nset_{i}]} \omega_{i,j}(t)$. This decision means that a packet from the link layer queue $V_{l^{*}}$ is chosen according to FIFO rule and transmitted. Note that this scheduling algorithm only performs intra-node scheduling, \ie it determines from which link layer queue packets should be transmitted, but it does not determine which node should transmit, which is handled by 802.11 MAC.

Diff-subMax has several nice features. We show in Appendix A that the deterministic version of Diff-subMax provides utility optimality for the networks with  pre-determined inter-node scheduling such as CSMA/CA. Furthermore, Diff-subMax reduces the complexity of the algorithm and overhead significantly. In particular, each node $i$ calculates and compares weights $\omega_{i,j}(t)$ for each neighbor node. Therefore, the complexity is linear with the number of (neighbor) nodes. The overhead is also significantly reduced; each node needs to know the queue size only of its one-hop away neighbors.

\subsection{wDiff-subMax}
wDiff-subMax is a heuristic designed as an extension of Diff-subMax for the scenarios that link layer operations and data exchange (between the network and link layers) are not possible due to Wi-Fi firmware or driver restrictions or may not be desired. Therefore, wDiff-subMax does not employ any scheduling mechanism, but only the routing and flow control. The flow control algorithm is the same as in Diff-Max. Yet, the routing algorithm is updated as explained next.

Per-flow queues as well as per-link queues are required in (\ref{eq_stoc:routing}) to make the routing decision. If per-link queues are not available at the network layer, the routing decision may not be efficient as there may be (uncontrolled) congestion in the link layer queues. In order to make the routing decisions efficiently, we propose a heuristic called wDiff-subMax. The main idea behind wDiff-subMax is to react to link layer congestion, while still implementing (\ref{eq_stoc:routing}). To achieve this, wDiff-subMax employs acknowledgement (ACK) mechanism and uses an additive increase/multiplicative decrease (AIMD) algorithm.

wDiff-subMax labels each packet with a timestamp at the network layer. When a packet is received by the next hop, an ACK packet, echoing the timestamp of the packet, is transmitted back to the previous hop. The network layer of the previous hop receives ACKs and determines round trip time (RTT) for each packet. $RTT_{i,j}^{s}(t_{i}^{'k})$ is the average round trip time of the ACKs received in the last slot (\ie at slot $t_{i}^{'k-1}$), and $RTT_{i,j}^{s}$ is the average round trip time of the packets.

wDiff-subMax keeps a windows size $W_{i,j}^{s}(t_{i}^{'k})$ for link $i-j$ and flow $s$ at slot $t_{i}^{'k}$. At each slot $t_{i}^{'k}$, the routing parameter $f_{i,j}^{s}(t_{i}^{'k})$ is set to $W_{i,j}^{s}(t_{i}^{'k})$ and $f_{i,j}^{s}(t_{i}^{'k})$ packets are passed to the link layer. wDiff-subMax determines the window size according to AIMD as explained next.

If $U_{i}^{s}(t_{i}^{'k}) - U_{j}^{s}(t_{i}^{'k}) > 0$ and $RTT_{i,j}^{s}(t_{i}^{'k}) < RTT_{i,j}^{s}$, then $W_{i,j}^{s}(t_{i}^{'k})$ is increased by 1. Note that, in this case, per-slot RTT is smaller than the average RTT, which means that congestion level in the link layer is low, and there is a positive queue backlog difference between the two nodes. Thus, more packets can be transmitted over this link, so $W_{i,j}^{s}(t_{i}^{'k})$ is increased.
On the other hand, if $U_{i}^{s}(t_{i}^{'k}) - U_{j}^{s}(t_{i}^{'k}) > 0$ and $RTT_{i,j}^{s}(t_{i}^{'k}) > RTT_{i,j}^{s}$, then $W_{i,j}^{s}(t_{i}^{'k})$ is decreased by 1 since the link layer congestion level is high, and less packets should be transmitted over this link. If none of the packets in the last slot is ACKed, this means that congestion is very high, and packets are dropped. In this case, $W_{i,j}^{s}(t_{i}^{'k})$ is halved so that the number of packets over link $i-j$ could be reduced sharply. After $W_{i,j}^{s}(t_{i}^{'k})$ is determined, $f_{i,j}^{s}(t_{i}^{'k})$ is set to $W_{i,j}^{s}(t_{i}^{'k})$ and $f_{i,j}^{s}(t_{i}^{'k})$ packets are passed to the link layer. Note that wDiff-subMax, similar to Diff-subMax, reduces computational complexity and overhead significantly as compared to Diff-Max.

\section{\label{sec:performance}Performance Evaluation}
\subsection{Numerical Simulations}
\begin{figure*}[t!]
\centering
\subfigure[Triangle topology]{ \scalebox{.6}{\includegraphics[bb=0 0 240 172]{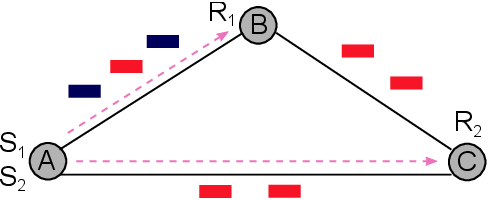}} }
\subfigure[Diamond topology]{ \scalebox{.6}{\includegraphics[bb=0 0 240 172]{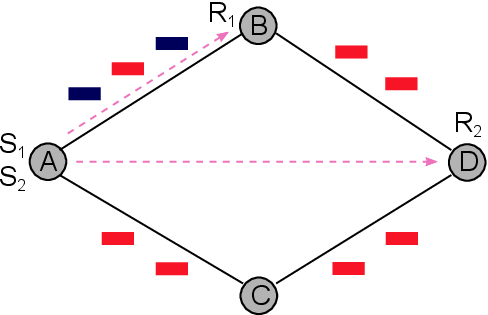}} }
\subfigure[Grid topology]{ \scalebox{.54}{\includegraphics[bb=0 0 291 172]{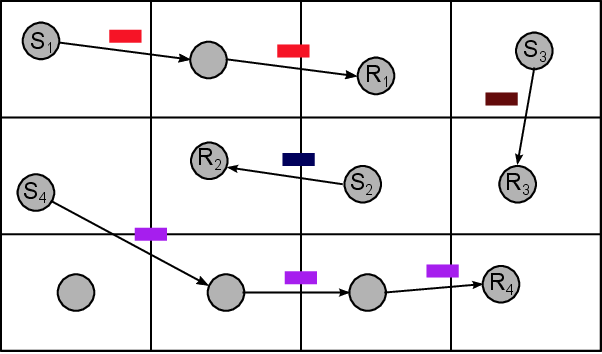}} }
\caption{Topologies used in simulations. (a) Triangle topology. There are two flows between sources; $S_1$, $S_2$ and receivers; $R_1$, $R_2$, \ie from node $A$ to $B$ ($S_1$ - $R_1$) and from node $A$ to $C$ ($S_2$ - $R_2$). (b) Diamond topology. There are two flows between sources; $S_1$, $S_2$ and receivers; $R_1$, $R_2$, \ie from node $A$ to $B$ ($S_1$ - $R_1$) and from node $A$ to $D$ ($S_2$ - $R_2$). (c) Grid topology. 12 nodes are randomly placed over $4 \times 3$ grid. An example node distribution and possible flows are illustrated in the figure.}
\label{fig:topologies}
\end{figure*}

We simulate Diff-Max as well as the standard backpressure in an idealized time slotted system. In particular, we consider triangle and diamond topologies shown in Figs.~\ref{fig:topologies}(a) and (b). In both topologies, there are two flows; $S_1-R_1$ and $S_2-R_2$, and all nodes are capable of forwarding packets to their neighbors. The simulation duration is $10000$ slots, and each simulation is repeated for 10 seeds. Each slot is in the $ON$ or $OFF$ state according to an i.i.d. random process with a given loss probability. The utility function in these simulations is $\log$ utility, \ie $g_s(x_s(t))=\log(x_s(t))$.

\begin{figure*}[t!]
\begin{center}
\subfigure[Throughput of $S_1-R_1$ flow.]{{\includegraphics[width=4.5cm]{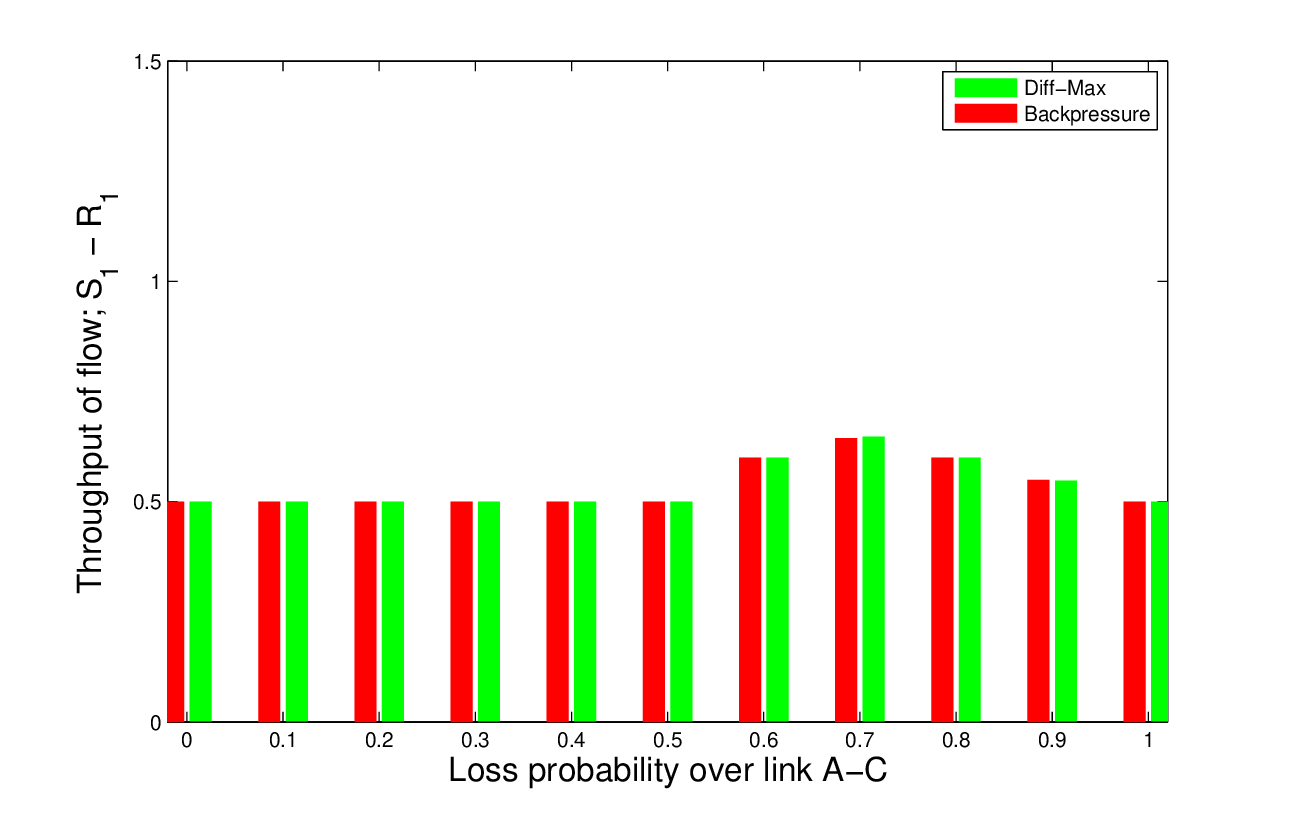}}}
\subfigure[Throughput of $S_2-R_2$ flow.]{{\includegraphics[width=4.5cm]{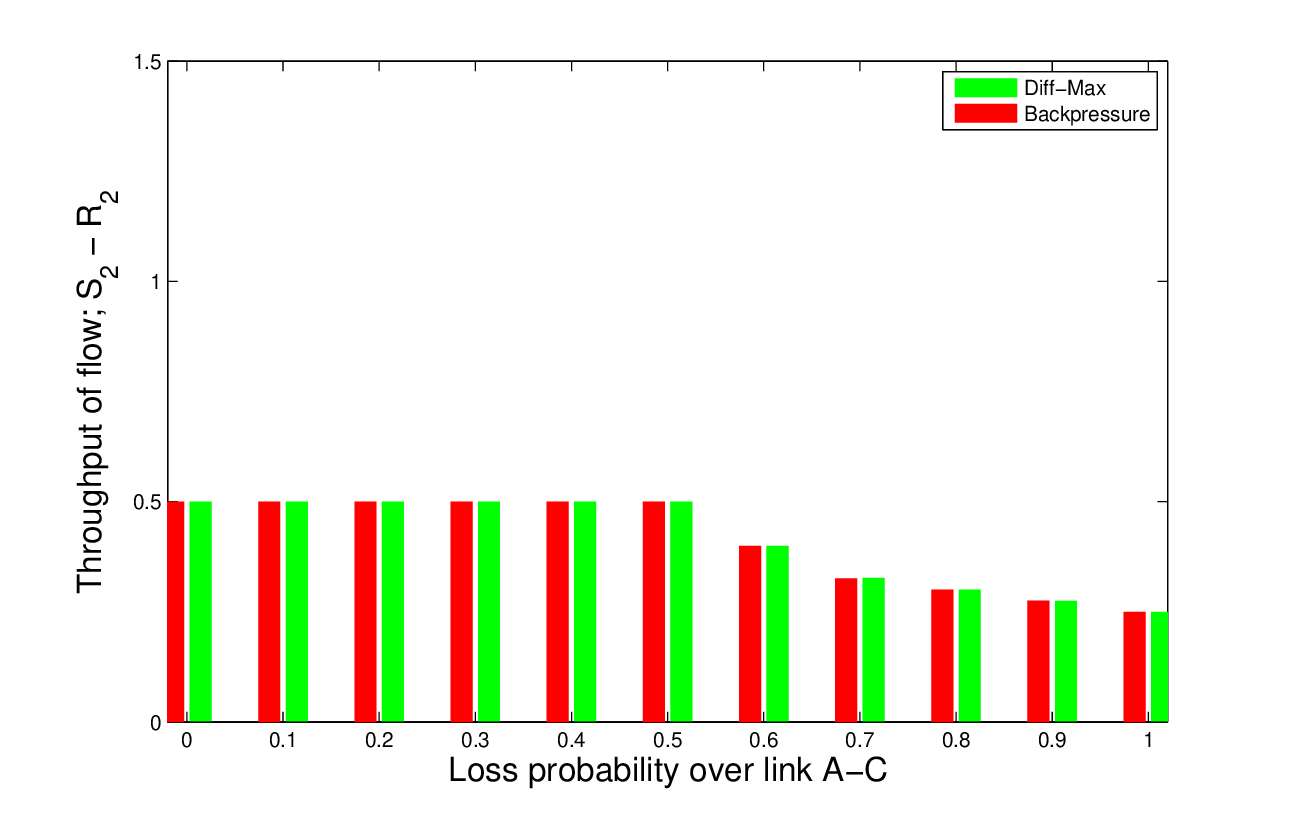}}}
\subfigure[Total utility.] {{\includegraphics[width=4.1cm]{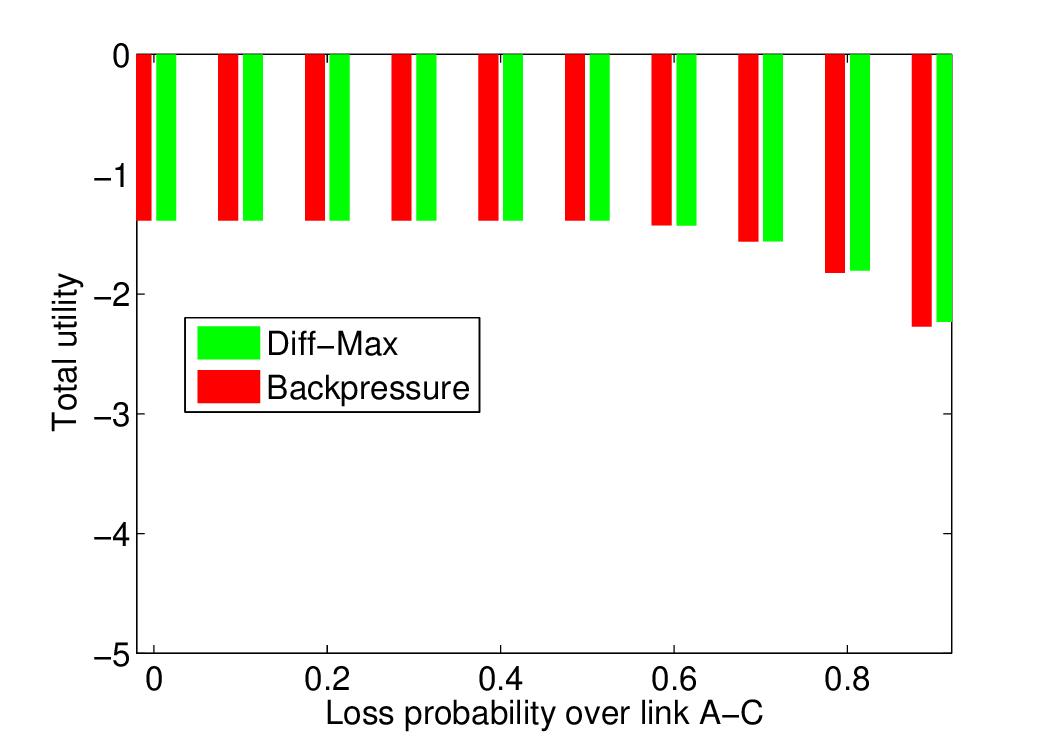}}}
\end{center}
\begin{center}
\vspace{-10pt}
\caption{\label{fig:matlab_results_11} Numerical results for the triangle topology shown in Fig.~\ref{fig:topologies}(a). The loss is over link $A-C$. (a) Throughput of $S_1-R_1$ flow. (b) Throughput of $S_2-R_2$ flow. (c) Total utility.}
\vspace{-10pt}
\end{center}
\end{figure*}

\begin{figure*}[t!]
\begin{center}
\subfigure[Throughput of $S_1-R_1$ flow.]{{\includegraphics[width=4.5cm]{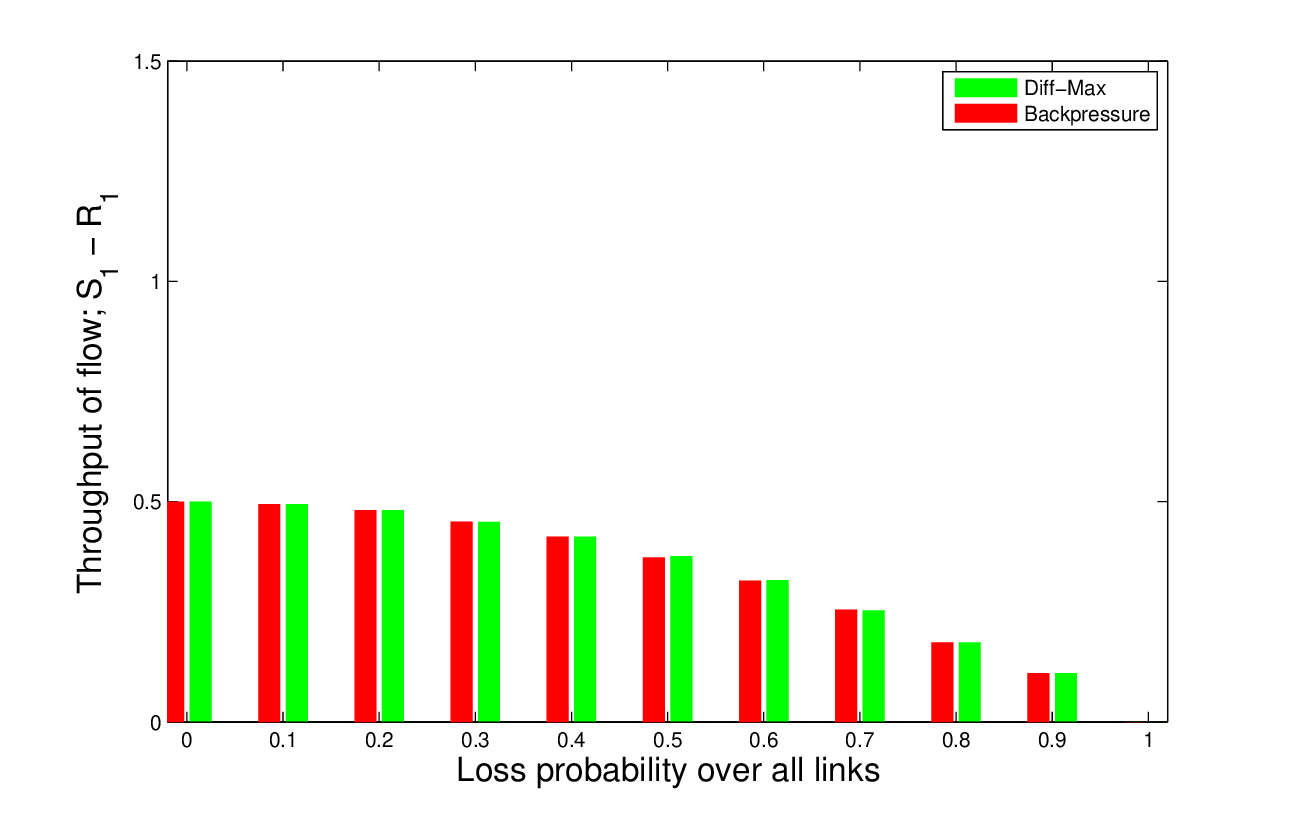}}}
\subfigure[Throughput of $S_2-R_2$ flow.]{{\includegraphics[width=4.5cm]{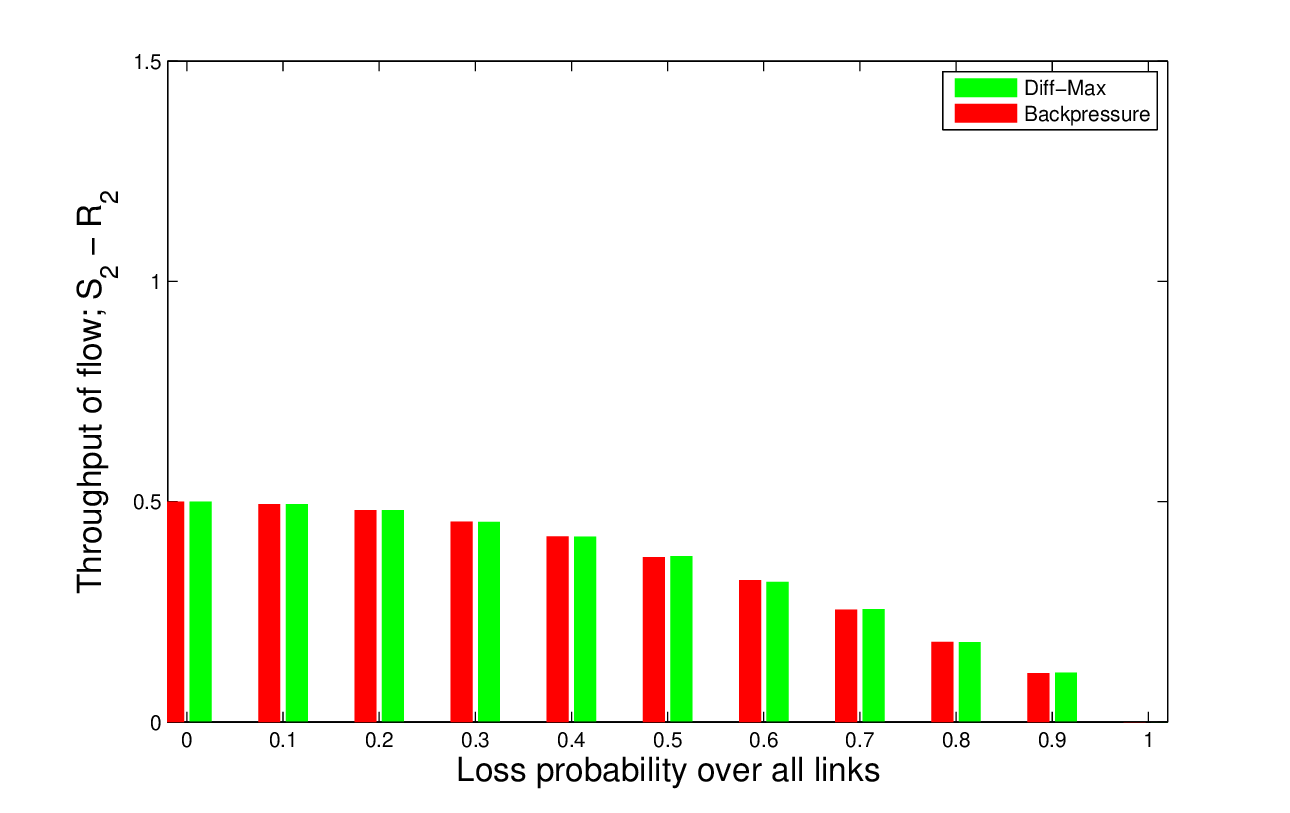}}}
\subfigure[Total utility.] {{\includegraphics[width=4.1cm]{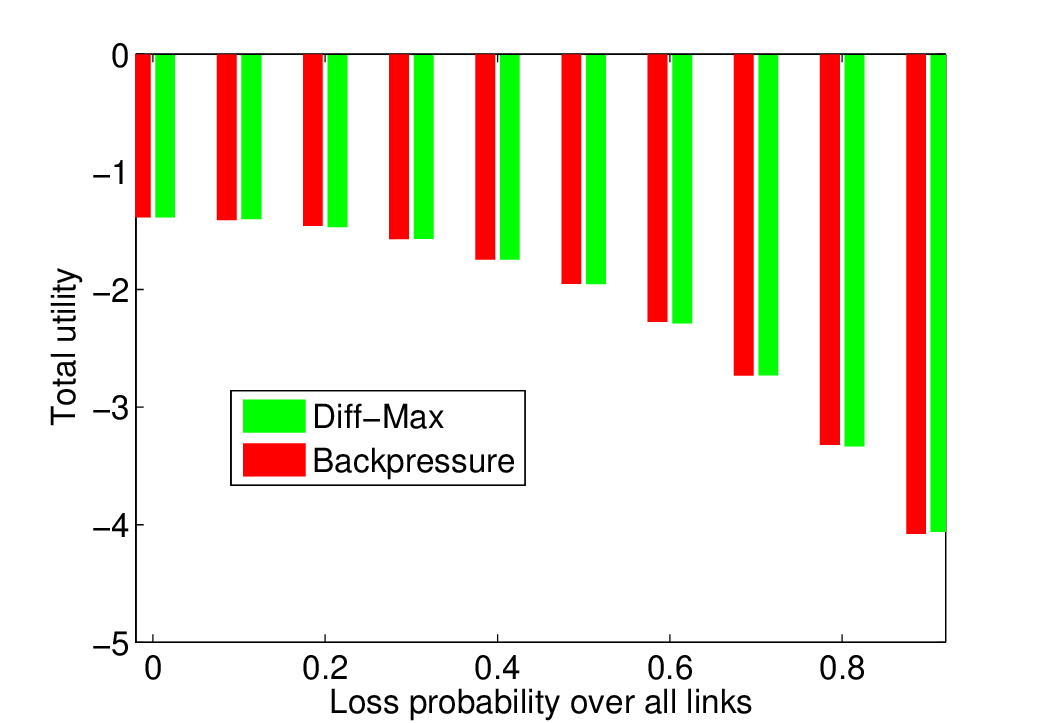}}}
\end{center}
\begin{center}
\vspace{-10pt}
\caption{\label{fig:matlab_results_12} Numerical results for the triangle topology shown in Fig.~\ref{fig:topologies}(a). The loss is over all links. (a) Throughput of $S_1-R_1$ flow. (b) Throughput of $S_2-R_2$ flow. (c) Total utility.}
\vspace{-10pt}
\end{center}
\end{figure*}

\begin{figure*}[t!]
\begin{center}
\subfigure[Throughput of $S_1-R_1$ flow.]{{\includegraphics[width=4.5cm]{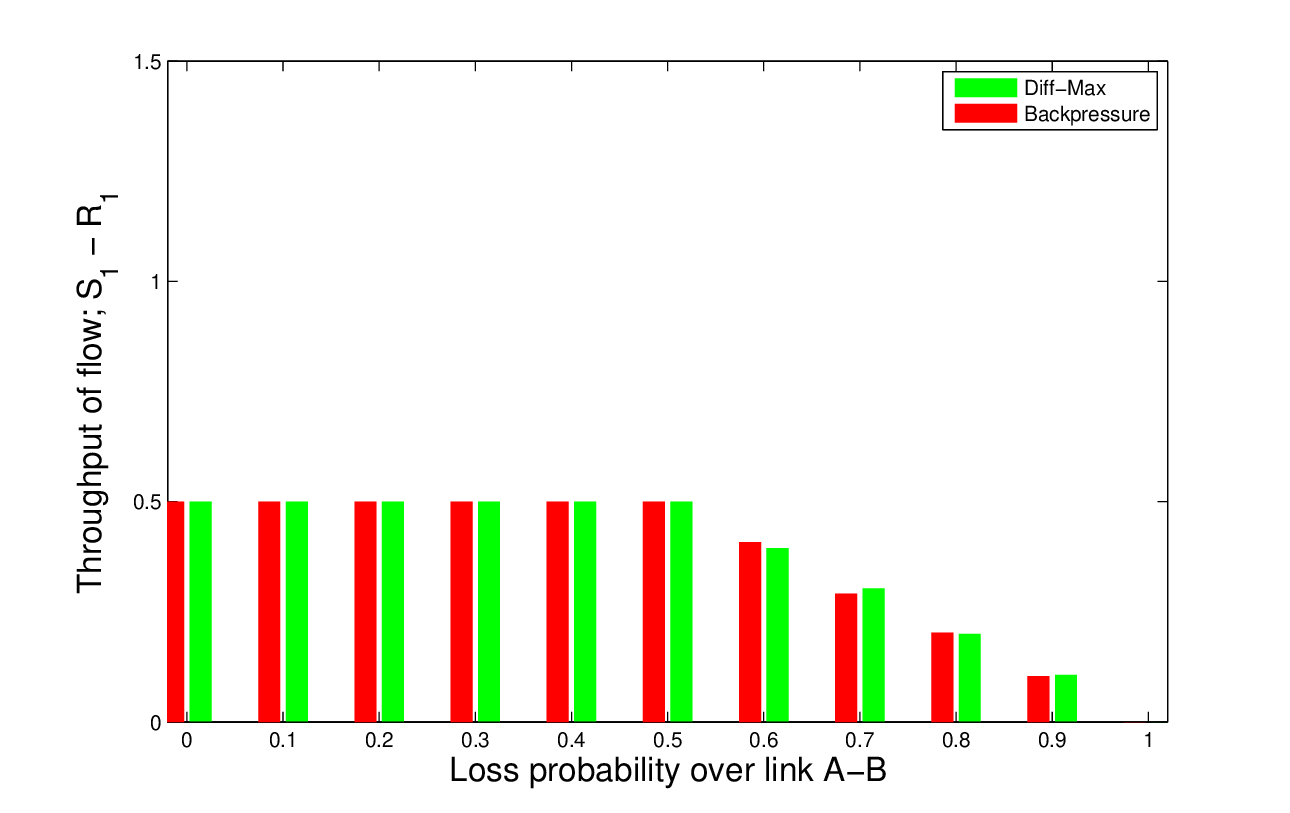}}}
\subfigure[Throughput of $S_2-R_2$ flow.]{{\includegraphics[width=4.5cm]{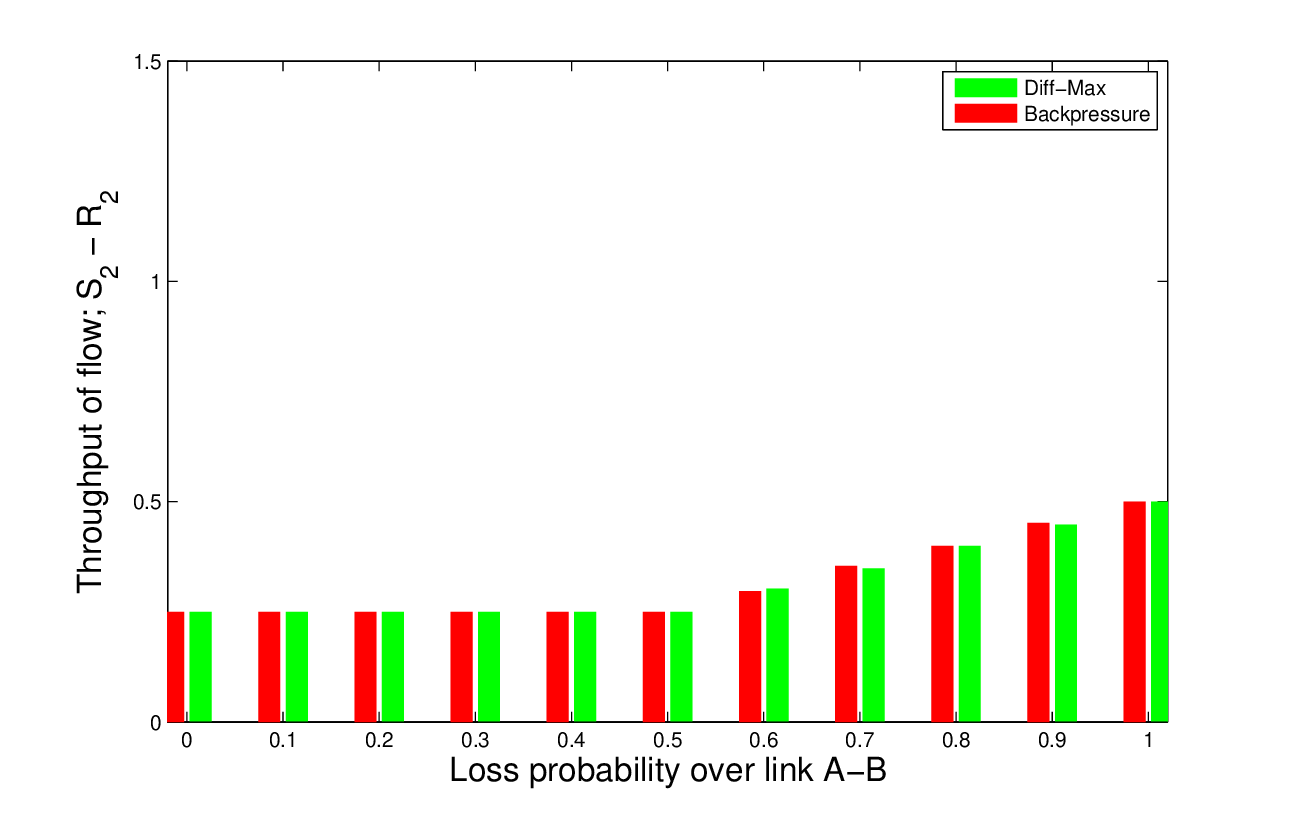}}}
\subfigure[Total utility.] {{\includegraphics[width=4.1cm]{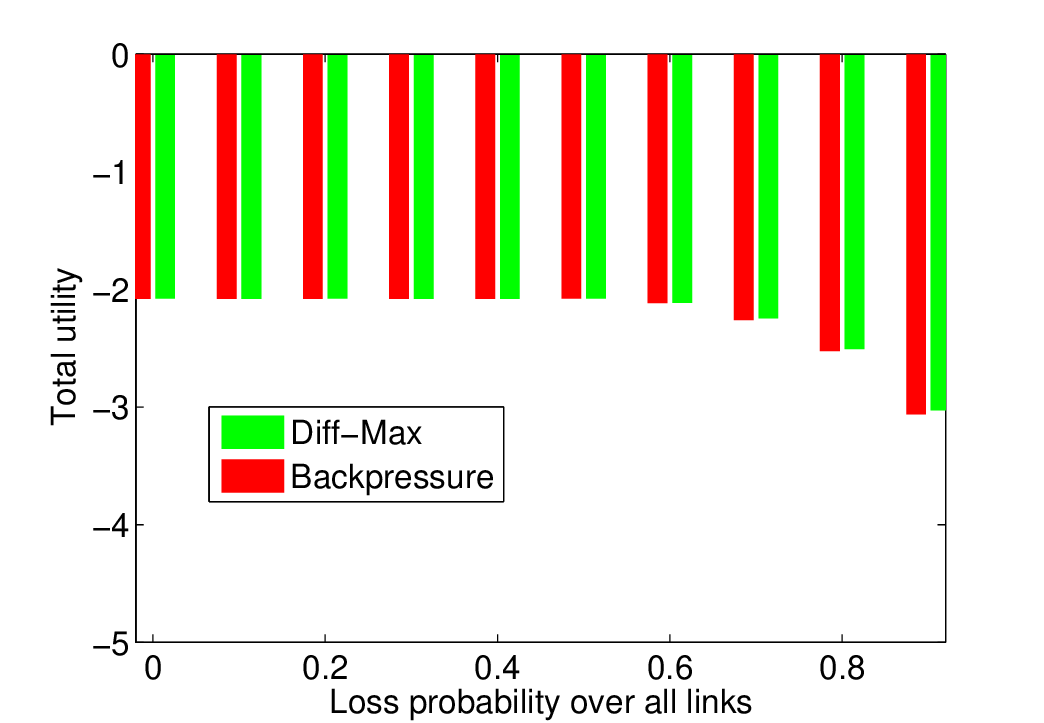}}}
\end{center}
\begin{center}
\vspace{-10pt}
\caption{\label{fig:matlab_results_21} Numerical results for the diamond topology shown in Fig.~\ref{fig:topologies}(b). The loss is over link $A-B$. (a) Throughput of $S_1-R_1$ flow. (b) Throughput of $S_2-R_2$ flow. (c) Total utility.}
\vspace{-10pt}
\end{center}
\end{figure*}

\begin{figure*}[t!]
\begin{center}
\subfigure[Throughput of $S_1-R_1$ flow.]{{\includegraphics[width=4.5cm]{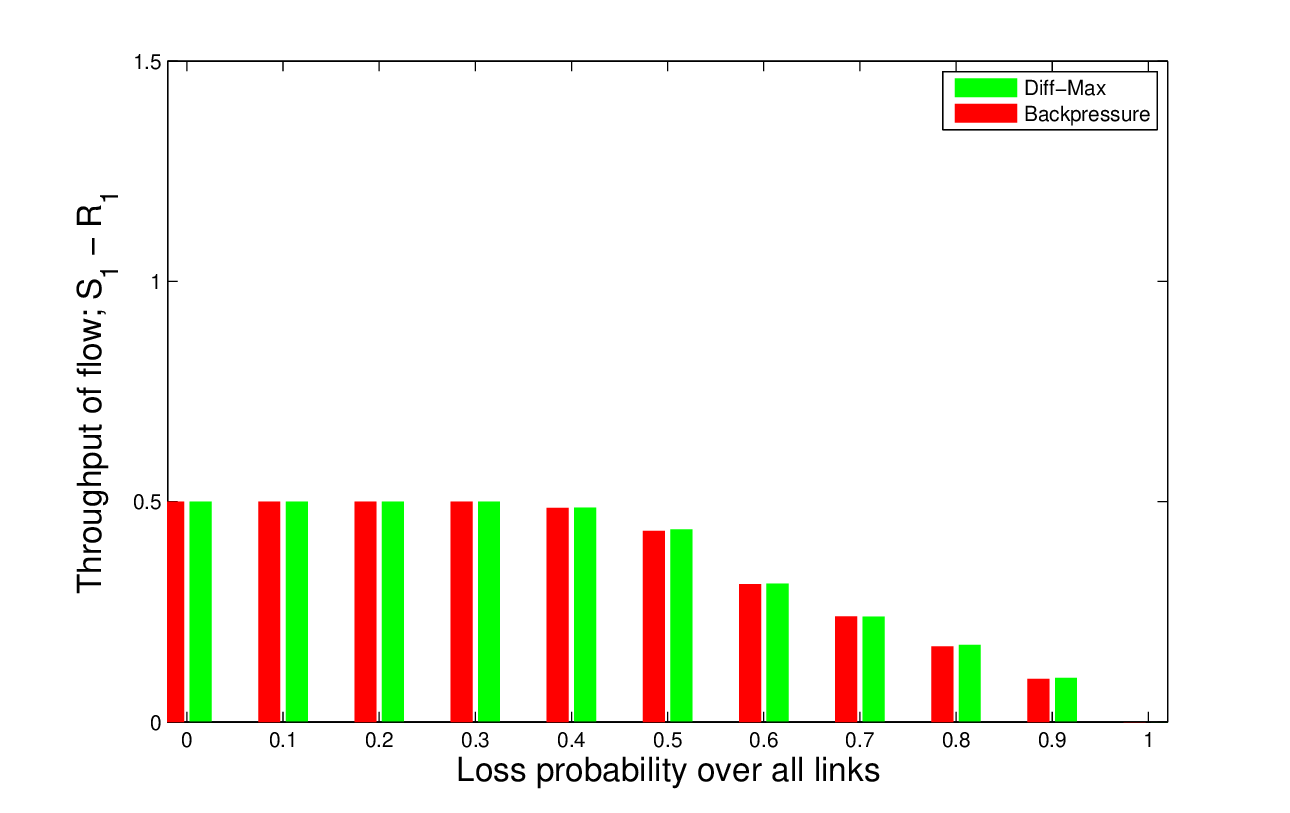}}}
\subfigure[Throughput of $S_2-R_2$ flow.]{{\includegraphics[width=4.5cm]{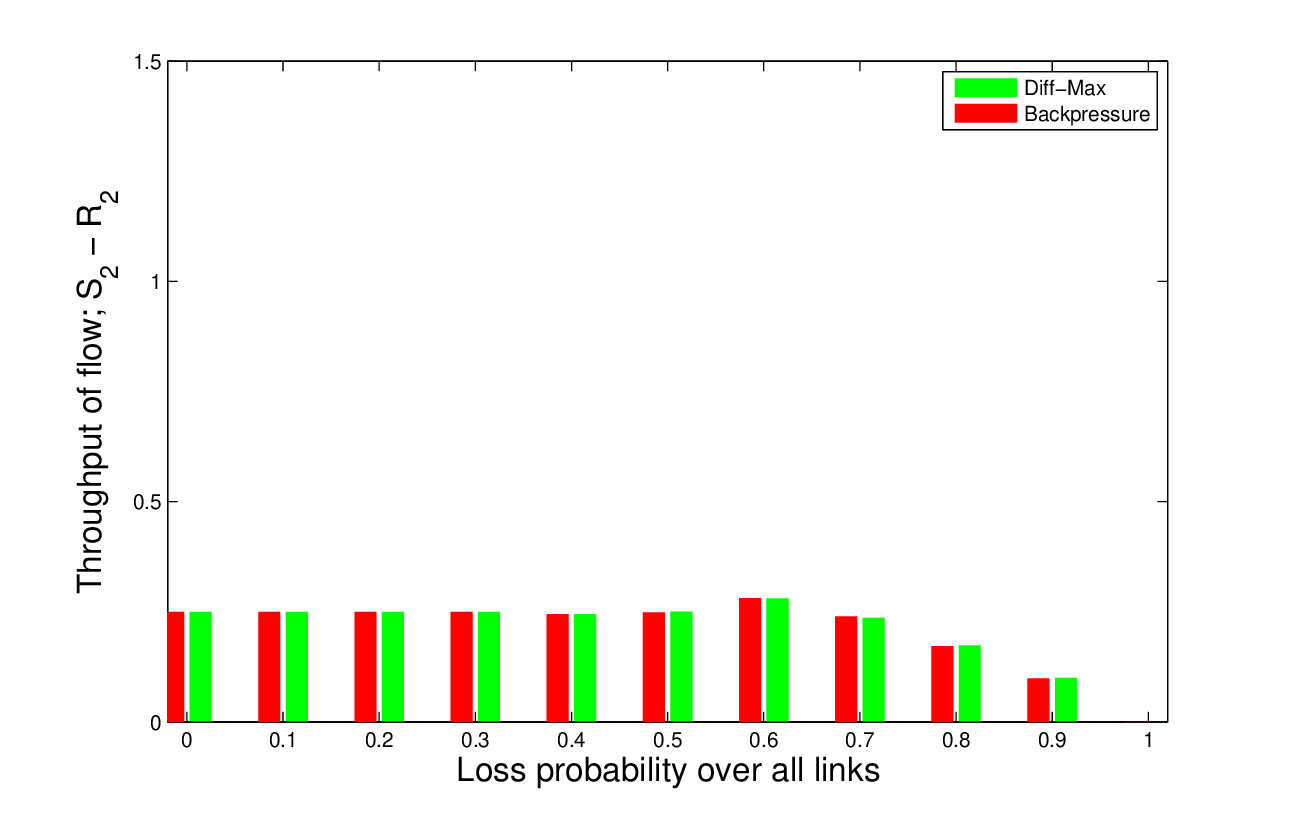}}}
\subfigure[Total utility.] {{\includegraphics[width=4.1cm]{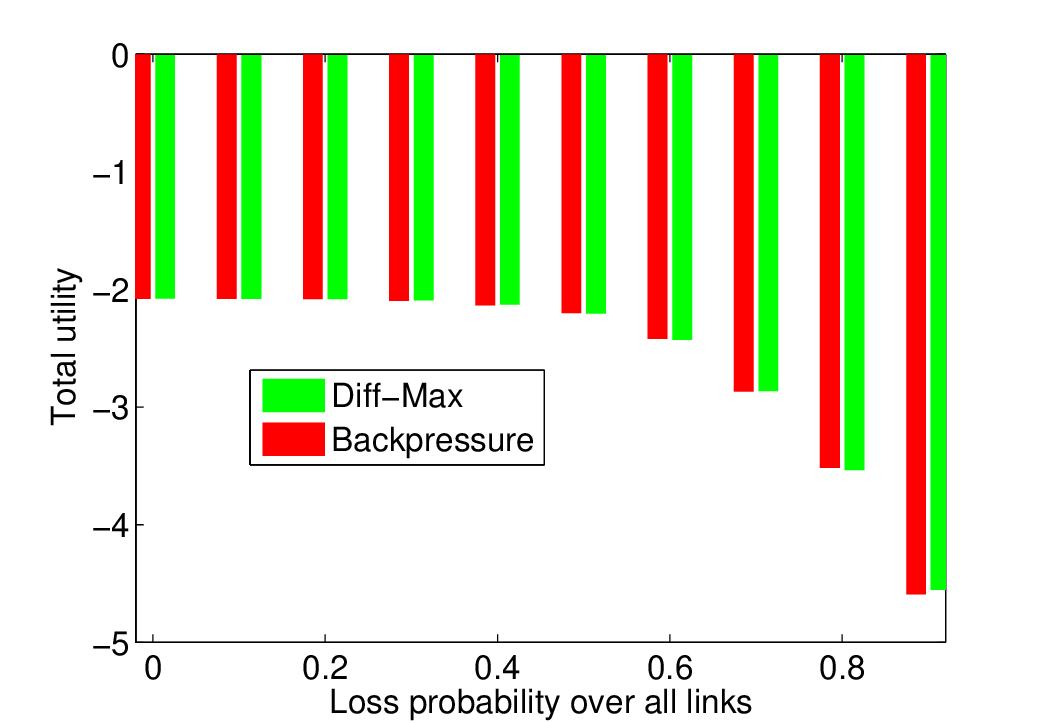}}}
\end{center}
\begin{center}
\vspace{-10pt}
\caption{\label{fig:matlab_results_22} Numerical results for the diamond topology shown in Fig.~\ref{fig:topologies}(b). The loss is over all links. (a) Throughput of $S_1-R_1$ flow. (b) Throughput of $S_2-R_2$ flow. (c) Total utility.}
\vspace{-10pt}
\end{center}
\end{figure*}

Fig.~\ref{fig:matlab_results_11} shows the throughput and total utility (aggregated over per-flow utilities) vs. the loss probability for the triangle topology when the link  $A-C$ is lossy. As can be seen, the throughput and the total utility of Diff-Max is equal to that of backpressure.  Similar results are observed in Fig.~\ref{fig:matlab_results_11} for the same setup when all links are lossy. Note that the total utilities in Fig.~\ref{fig:matlab_results_11} and Fig.~\ref{fig:matlab_results_12} are negative as we employ $\log$ utility in these simulations. 
As can be seen, the throughput and utility of Diff-Max is equal to that of backpressure. The same results are shown for the diamond topology in Fig.~\ref{fig:matlab_results_21} and Fig.~\ref{fig:matlab_results_22}. These results show that Diff-Max achieves the same throughput and utility as backpressure.

\subsection{ns-2 Simulations}
In this section, we simulate our schemes, Diff-Max, Diff-subMax, wDiff-subMax using the ns-2 simulator \cite{ns2}. The simulation results show that our schemes significantly improve throughput, utility, and delay performance as compared to Ad hoc On-Demand Distance Vector (AODV) \cite{aodv}, and Destination-Sequenced Distance-Vector Routing (DSDV) \cite{dsdv} routing schemes. Next, we present the simulator setup and results in detail.

\subsubsection{Setup}
We considered three topologies: the diamond topology shown in Fig.~\ref{fig:topologies}(b); a grid topology shown in Fig.~\ref{fig:topologies}(c), and a random topology. In the diamond topology, the nodes are placed over $500m \times 500m$ terrain. Two flows are transmitted from node $A$ to nodes $B$ and $D$. In the grid topology, $4 \times 3$ cells are placed over a $800m \times 600m$ terrain; $12$ nodes are randomly placed in the cells. In the grid topology, each node can communicate with other nodes in its cells or with the ones in neighboring cells. Four flows are generated randomly. In the random topology, 20 nodes are randomly generated and located over a $800m \times 800m$ terrain according to a uniform distribution. Ten flows are generated, and transmitted between randomly selected nodes.

We consider CBR flows, which start at random times within the first $5sec$ and remain on until the end of the simulation which is $100sec$. The CBR flows generate packets with inter-arrival times $0.01ms$. IEEE 802.11 is used in the MAC layer (with updates for Diff-Max implementation as explained in Section~\ref{sec:algs}). We simulated a Rayleigh fading channel with average channel loss rates $0, 20, 30, 40, 50\%$.\footnote{We consider the loss rates in the range up to $50\%$, because previous studies of IEEE 802.11b based wireless mesh networks \cite{MPLOT_ref1}, \cite{MPLOT_ref2}, have reported packet loss rates as high as 50\%.} We have repeated each $100sec$ simulation for $10$ seeds.

The channel capacity is $1Mbps$,  the buffer size at each node is set to $1000$ packets, packet sizes are set to $1000B$. We compare our schemes; Diff-Max, Diff-subMax, and wDiff-subMax with AODV and DSDV in terms of transport-level throughput, total utility (added over all flows), and average delay (averaged over all packets and flows). We employ $\log$ utility in our simulations, \ie $g_s(x_s(t))=\log(x_s(t))$. On the other hand, packet delay is measured at the transport layer. Let $r_{s,k}$ be the time that the $k$th packet of flow $s$ is received by the transport layer at the receiver side, and $t_{s,k}$ be the time that the same packet is seen by the transport layer at the transmitter side. Then, the packet delay is $r_{s,k} - t_{s,k}$. 

The Diff-Max parameters are set as follows. For the flow control algorithm; $T_i = 80ms$, $R_{i}^{max} = 20$ packets, $M = 200$. For the routing algorithm; $T_{i}^{'} = 10ms$, $F_{i}^{max}=4$ packets.

\subsubsection{Results}
\begin{figure}[t!]
\begin{center}
\subfigure[Total throughput.]{{\includegraphics[width=5.8cm]{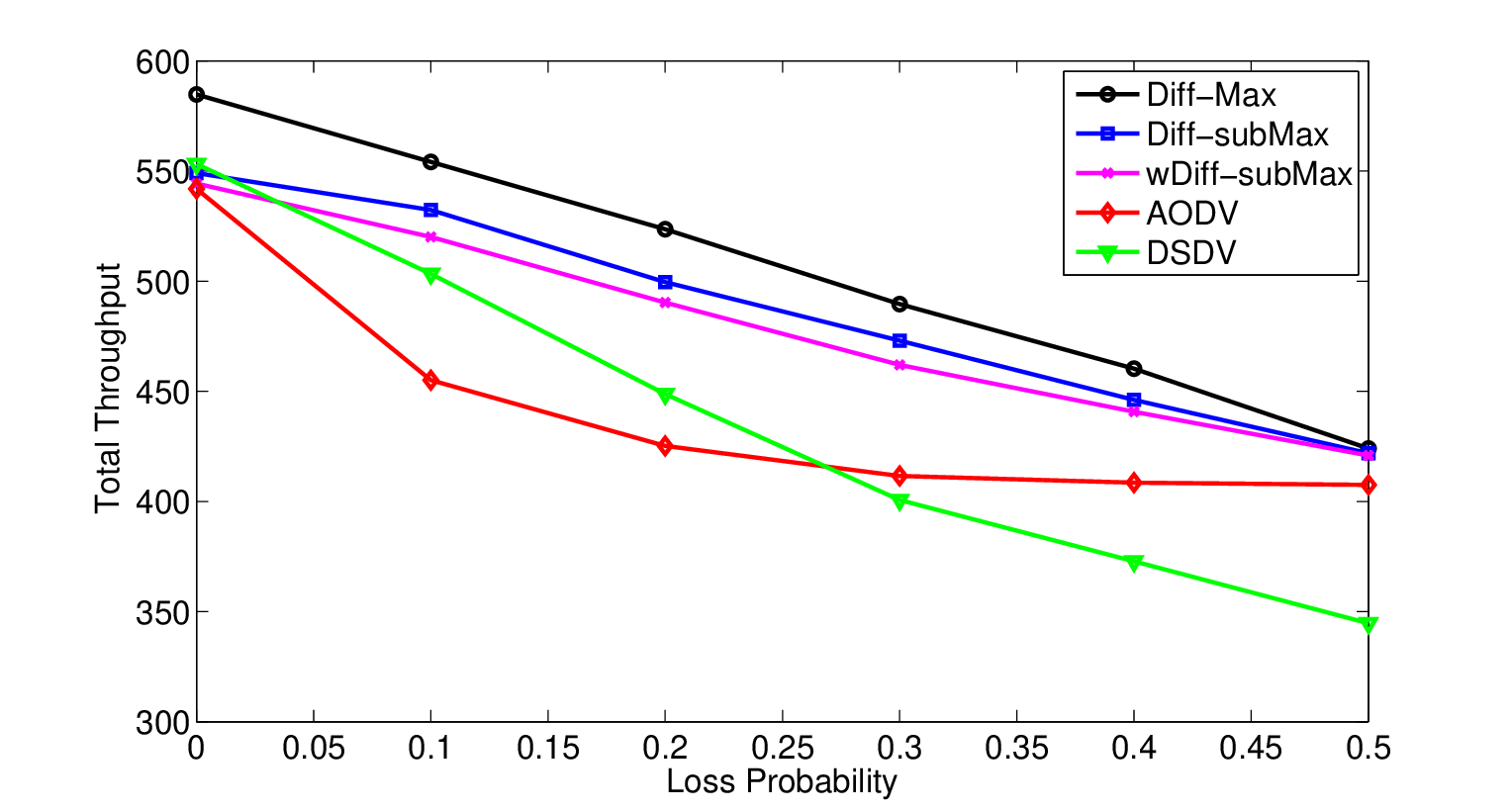}}}
\subfigure[Per-flow throughput.]{{\includegraphics[width=5.8cm]{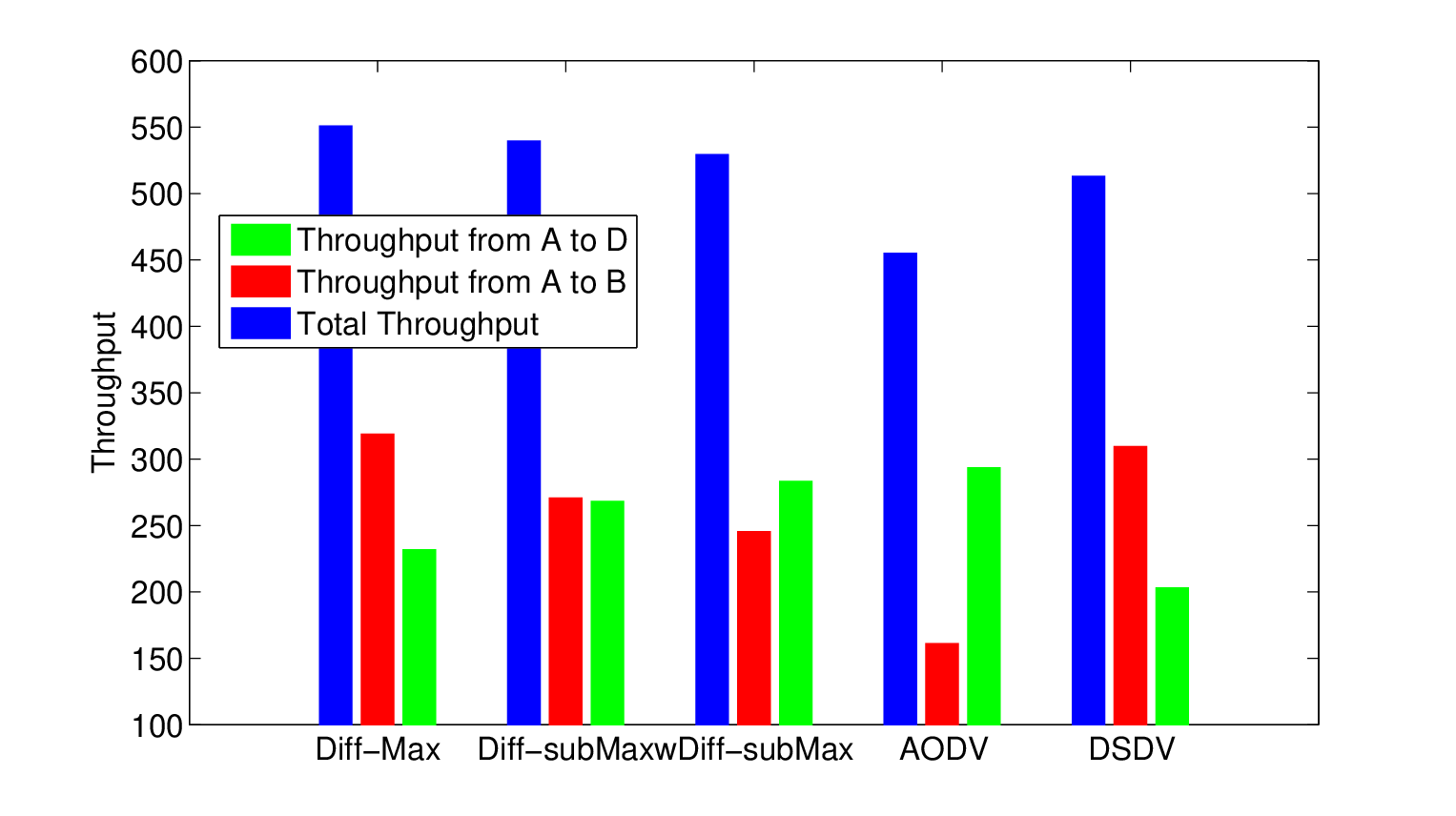}}}
\end{center}
\begin{center}
\vspace{-10pt}
\caption{\label{fig:ns2_results_diamond_1} Diamond topology. (a) Total throughput (in kbps) vs. average loss rate in the diamond topology. (b) Per-flow (as well as total) throughput of different policies when the average loss rate is set to $10\%$.}
\vspace{-10pt}
\end{center}
\end{figure}

Fig.~\ref{fig:ns2_results_diamond_1}(a) shows the simulation results for the diamond topology, where only link $A-B$ is lossy. Diff-Max performs better than the other schemes for the range of loss rates, since Diff-Max activates the links based on the per-link queue backlogs, loss rates, and link rates. On the other hand, Diff-subMax, wDiff-subMax, AODV, and DSDV use classical 802.11 MAC. When the loss rate over link $A-B$ increases, the total throughput of all the schemes decreases as expected. As can be seen, the decrease in our schemes; Diff-Max, Diff-subMax, wDiff-subMax is linear, while the decrease of AODV is quite sharp. The reason is that when AODV experiences loss over a path, it deletes the path and re-calculates new routes. Therefore, AODV does not transmit over lossy links for some time period and tries to find new routes, which reduces throughput. On the other hand, DSDV performs better than AODV at low loss rates thanks to keeping track of multiple routes and exploiting a new route when one becomes lossy. Yet, it is worse than AODV at higher loss rates, as it requires more packet exchanges among nodes at high loss rates, which consumes higher bandwidth and reduces throughput. Diff-subMax and wDiff-subMax improve throughput significantly as compared to both AODV and DSDV thanks to exploring routes to improve throughput. The improvement of our schemes is up to $22\%$ and $21\%$ over AODV and DSDV, respectively. Also, Diff-subMax and wDiff-subMax have similar throughput performance, which emphasizes the benefit of the routing part and the effective link layer queue estimation mechanism of wDiff-subMax.

Fig.~\ref{fig:ns2_results_diamond_1}(a) also shows that when loss rate is $50\%$, the throughput improvement of all schemes (except DSDV) are similar, because at $50\%$ loss rate, link $A-B$ becomes very inefficient, and all of the schemes transmit packets mostly from flow $A$ to $D$ over path $A-C-D$ and have similar performance at high loss rates. DSDV is worse because it requires more packet exchanges to keep the routing table, which wastes resources.

Fig.~\ref{fig:ns2_results_diamond_1}(b) elaborates more on the above discussion. It shows the throughput of two flows $A$ to $B$ and $A$ to $D$ as well as their total value when the loss rate is $10\%$ over link $A-B$. As can be seen, the rate of flow $A-B$ is very low in AODV as compared to our schemes, because AODV considers link $A-B$ to be broken at some periods during the simulation, while our schemes continue to transmit over this link. Although DSDV outperforms AODV, our schemes are still better than it in terms of throughput thanks to exploring routes to improve throughput.

\begin{figure}[t!]
\begin{center}
\subfigure[Average delay.]{{\includegraphics[width=5.8cm]{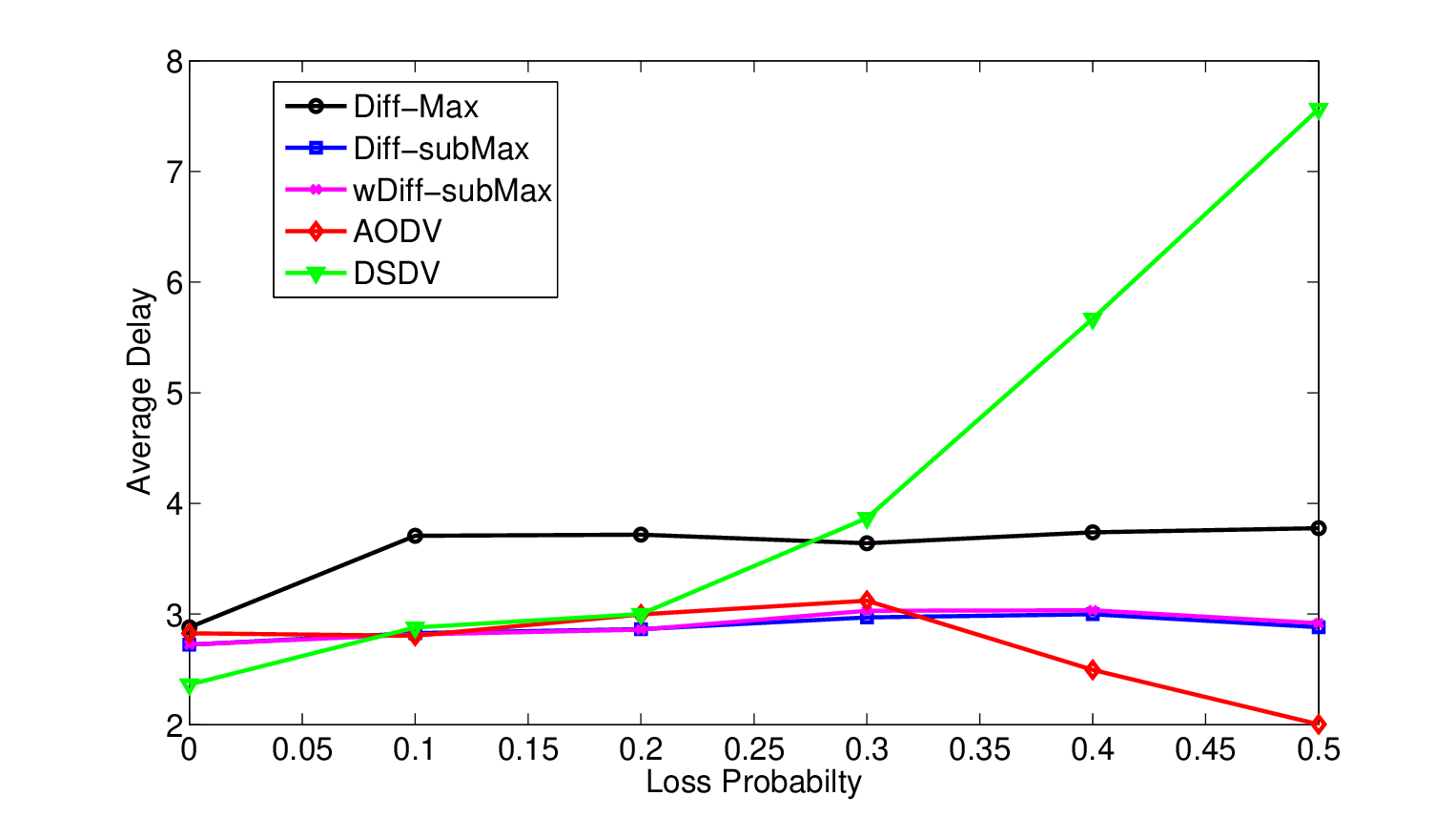}}} 
\subfigure[Per-flow delay.]{{\includegraphics[width=5.8cm]{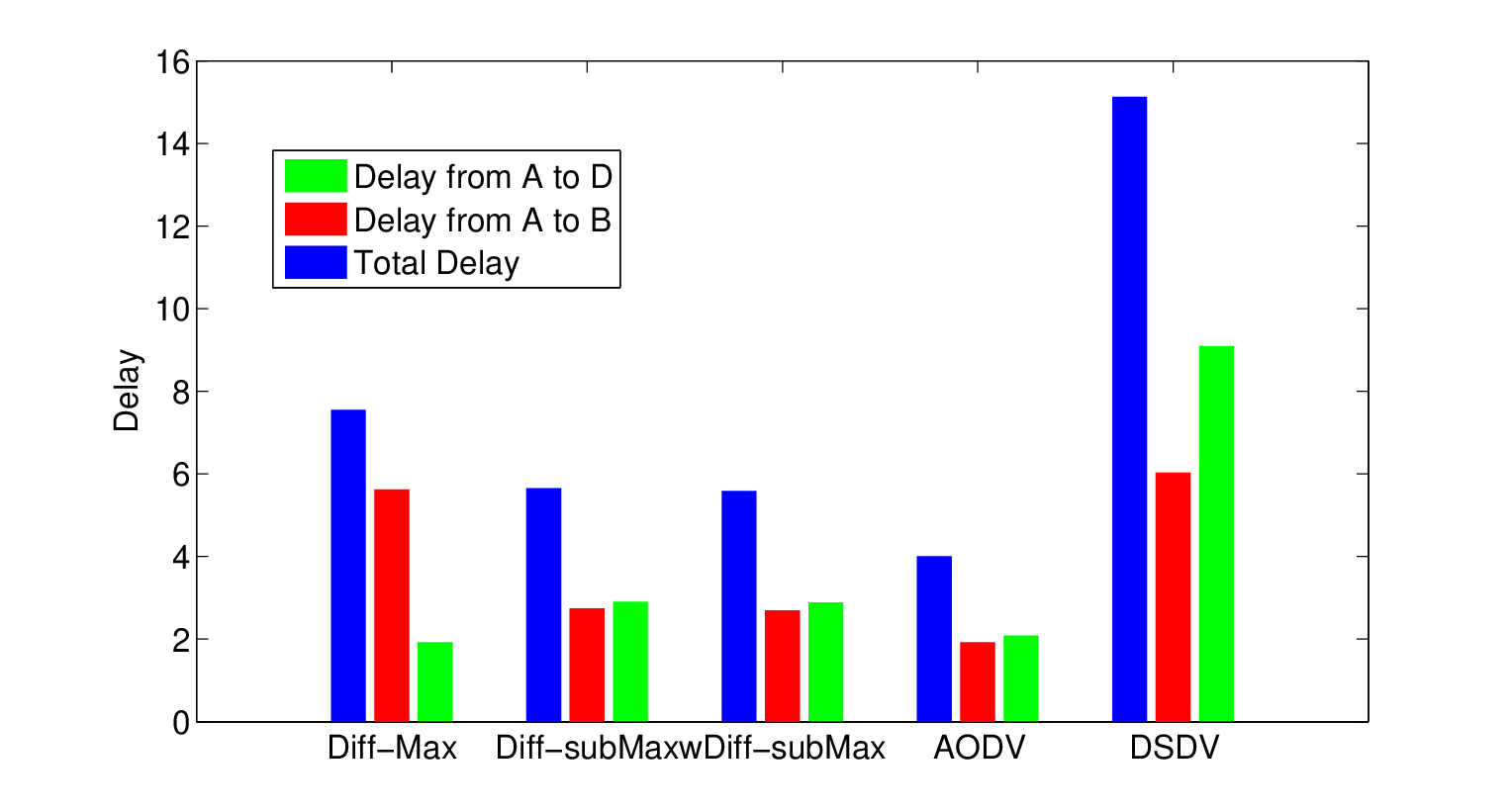}}} 
\end{center}
\begin{center}
\vspace{-10pt}
\caption{\label{fig:ns2_results_diamond_delay} Diamond topology. (a) Average delay (in sec) vs. average loss rate in the diamond topology. (b) Per-flow (as well as total) delay of different policies when the average loss rate is set to $50\%$.}
\vspace{-10pt}
\end{center}
\end{figure}

Fig.~\ref{fig:ns2_results_diamond_delay}(a) shows the delay vs. loss probability for the diamond topology, where only the link $A-B$ is lossy. As can be seen, Diff-Max introduces higher delay as compared to Diff-subMax and wDiff-subMax, because Diff-Max can delay packet transmission from some queues depending on their occupancy. In other words, Diff-Max transmits packets from the nodes with larger queue size, which may delay some packets significantly. On the other hand, Diff-subMax and wDiff-subMax transmit packets from the link layer queues based on 802.11 MAC scheduling, which reduces delay. On the other hand, the delay performance of Diff-subMax and wDiff-subMax is comparable to and better than AODV and DSDV for all loss rates, which shows that our algorithms are quite efficient in terms of delay. The delay performance of DSDV is high and increases with loss rate as DSDV should update its routing table periodically and needed, which increases packet delay \cite{dsdv}. Fig.~\ref{fig:ns2_results_diamond_delay}(b) shows per-flow and total delay of each algorithm when the loss rate over link $A-B$ is $50\%$. As can be seen, the delay of each flow is very large in DSDV while the delay performances of other algorithms are comparable.

\begin{figure}[t!]
\begin{center}
\subfigure[Total throughput.]{{\includegraphics[width=5.0cm]{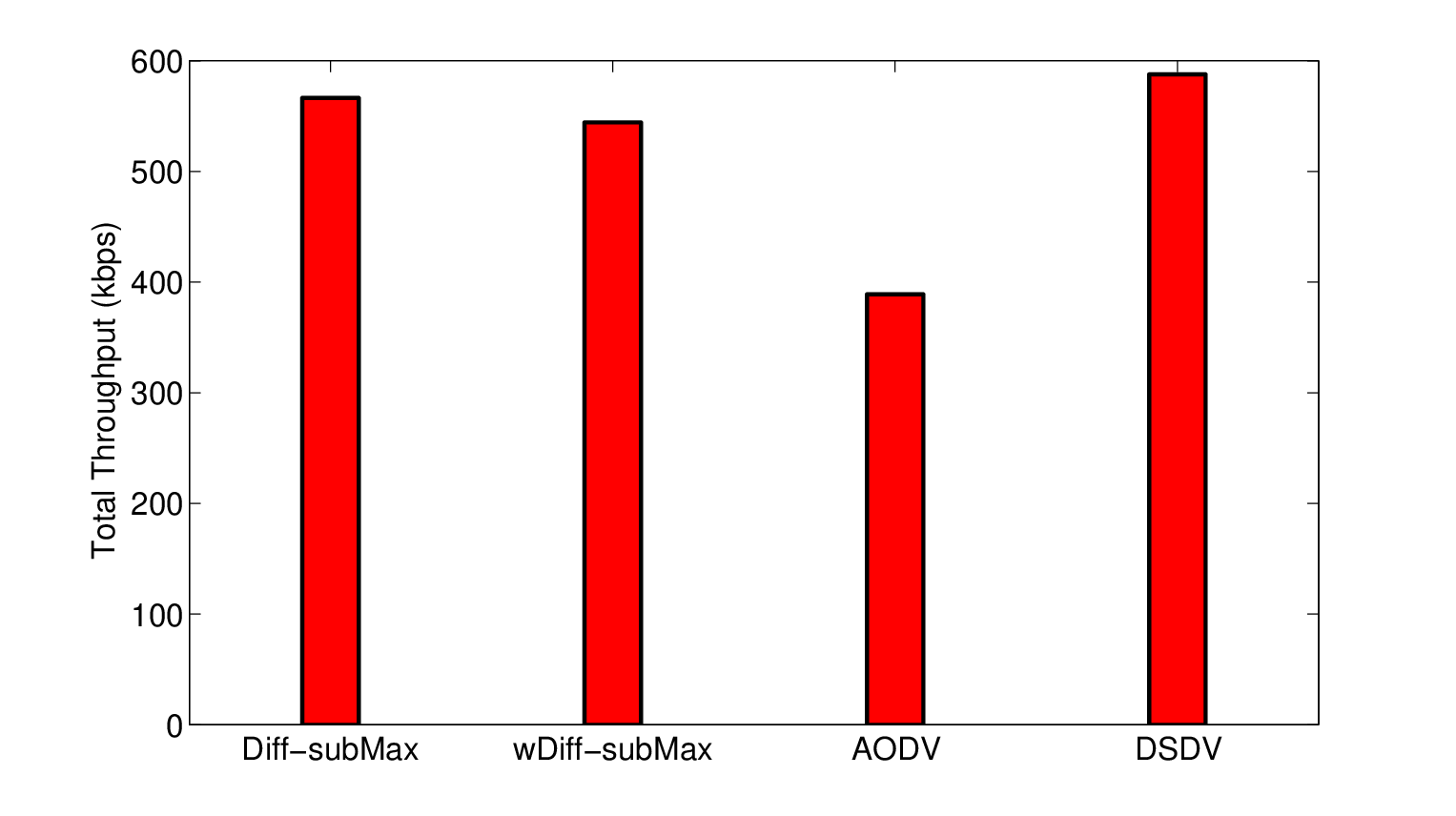}}} 
\subfigure[Total utility.]{{\includegraphics[width=5.0cm]{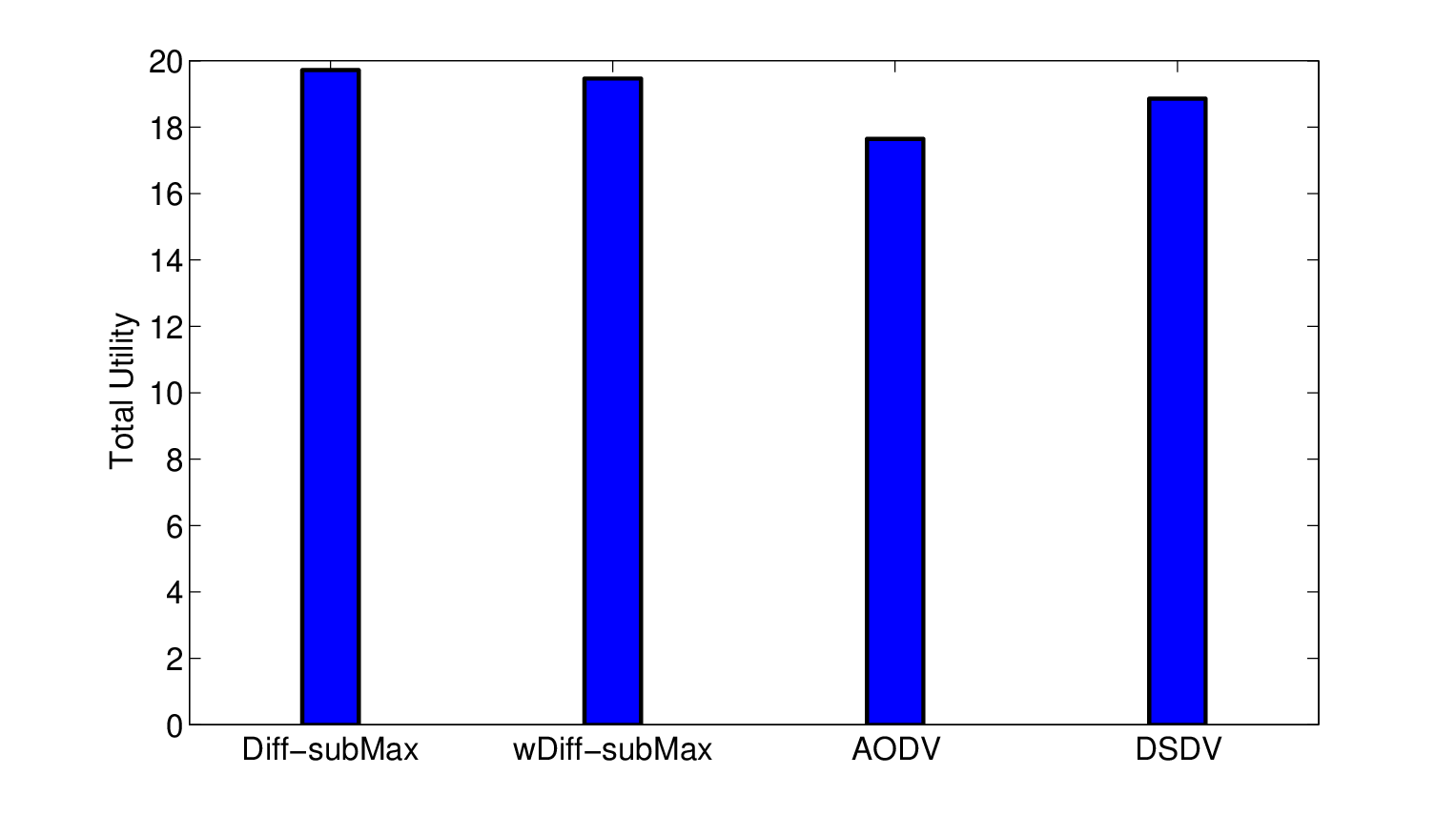}}} 
\end{center}
\begin{center}
\vspace{-10pt}
\caption{\label{fig:grid_top_res1} Grid Topology. (a) Total throughput, (b) total utility of Diff-subMax, wDiff-subMax, AODV, and DSDV. There is no loss over the links.}
\vspace{-10pt}
\end{center}
\end{figure}

Fig.~\ref{fig:grid_top_res1} presents the results for the grid topology. In this scenario, one third of the links, which are chosen randomly, are lossy with $10\%$ loss rate. Fig.~\ref{fig:grid_top_res1}(a) shows the total throughput of our schemes as well as AODV and DSDV.
Although the throughput performances of our schemes are better than AODV, the total throughput of DSDV is slightly better than our schemes. The reason is that DSDV treats some flows (with longer paths) unfairly, and they do not get much (or even any) opportunity to transmit. Since the flows with shorter paths can transmit most of the time, the total throughput of DSDV becomes better. On the other hand,
Fig.~\ref{fig:grid_top_res1}(b) shows that the total utilities of Diff-subMax and wDiff-subMax are better than DSDV. It is expected as our schemes are designed to maximize the total utility in (\ref{eq_stoc:rate_control}). In other words, even though the total throughput of DSDV may be higher at some scenarios, the total utility, which we maximize, of Diff-subMax and wDiff-subMax is higher.

\begin{figure}[t!]
\begin{center}
\subfigure[Utility.]{{\includegraphics[width=5.8cm]{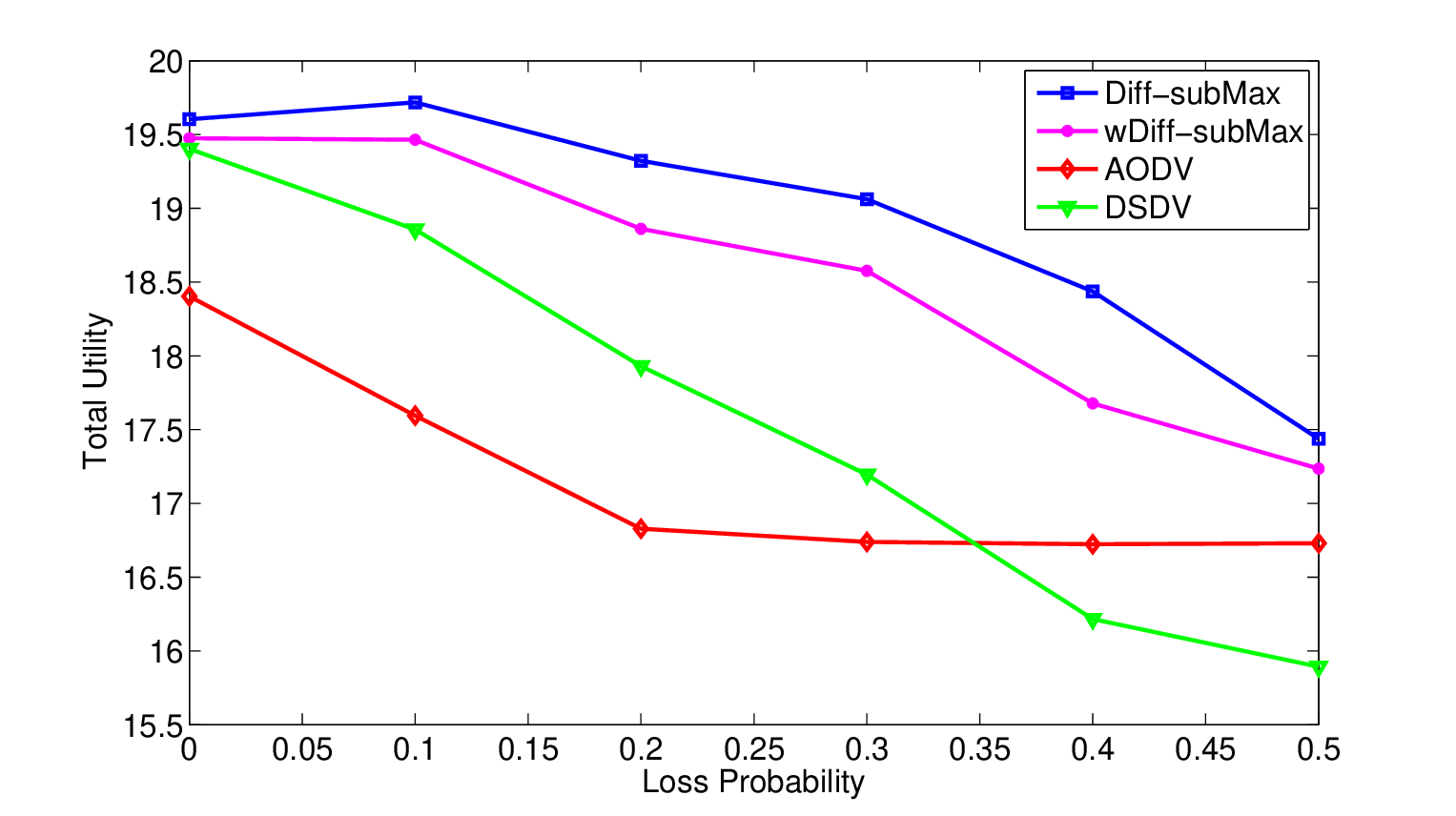}}} 
\subfigure[Delay.]{{\includegraphics[width=5.8cm]{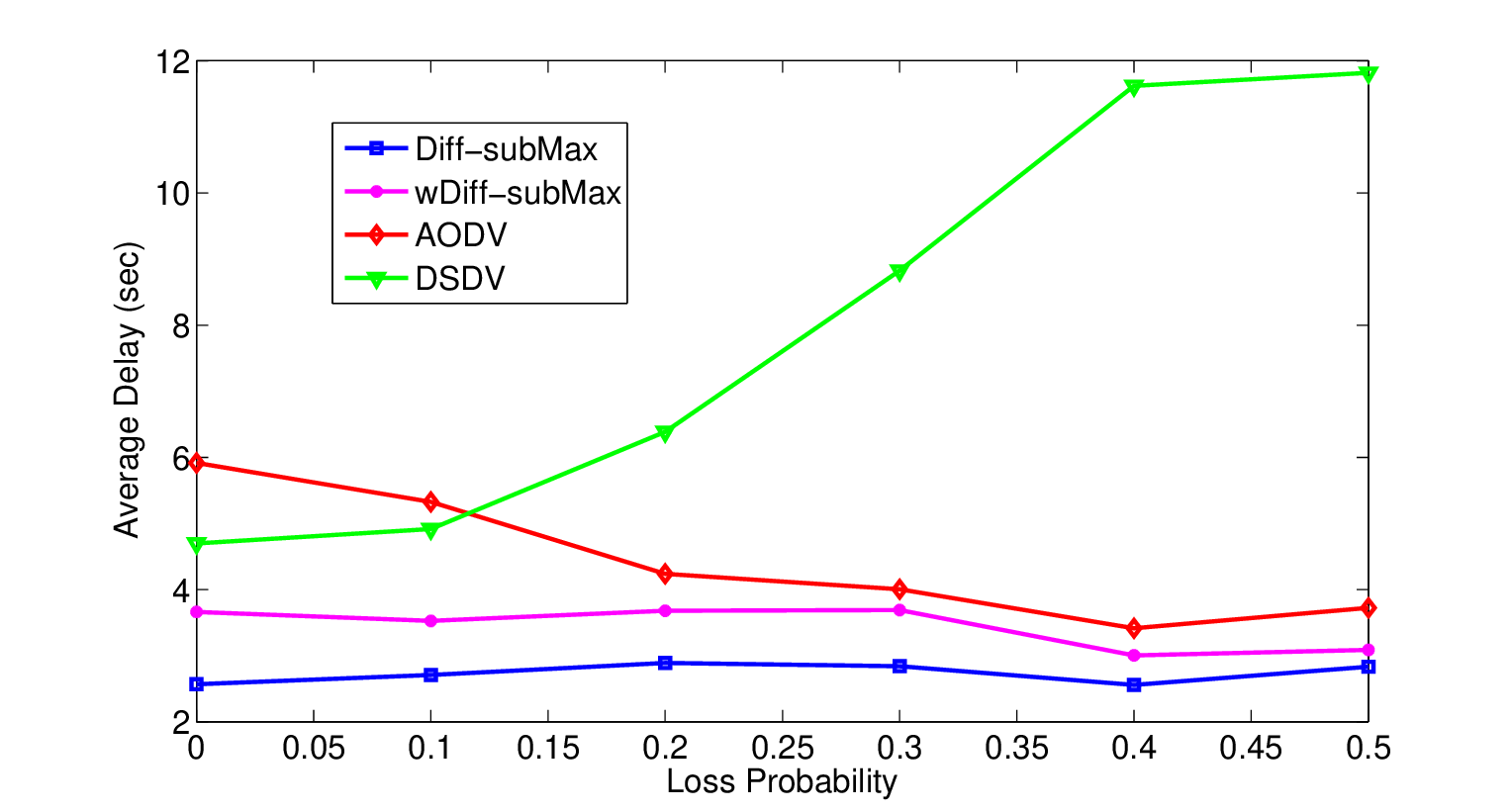}}} 
\end{center}
\begin{center}
\vspace{-10pt}
\caption{\label{fig:grid_top_res2} Grid Topology. (a) Total utility vs. average loss rate. (b) Average delay vs. average loss rate.}
\vspace{-10pt}
\end{center}
\end{figure}

Fig.~\ref{fig:grid_top_res2} presents the simulation results for the same grid topology setup. Fig.~\ref{fig:grid_top_res2}(a) shows the total utility vs. average loss rate for our schemes as well as AODV and DSDV. Our schemes Diff-subMax and wDiff-subMax significantly improve the total utility as compared to both AODV and DSDV. Fig.~\ref{fig:grid_top_res2}(b) considers the same setup, and presents average delay vs. average loss rate. As can be seen, Diff-subMax and wDiff-subMax significantly improves delay as compared to AODV and DSDV.

\begin{figure}[t!]
\begin{center}
\subfigure[Utility.]{{\includegraphics[width=5.8cm]{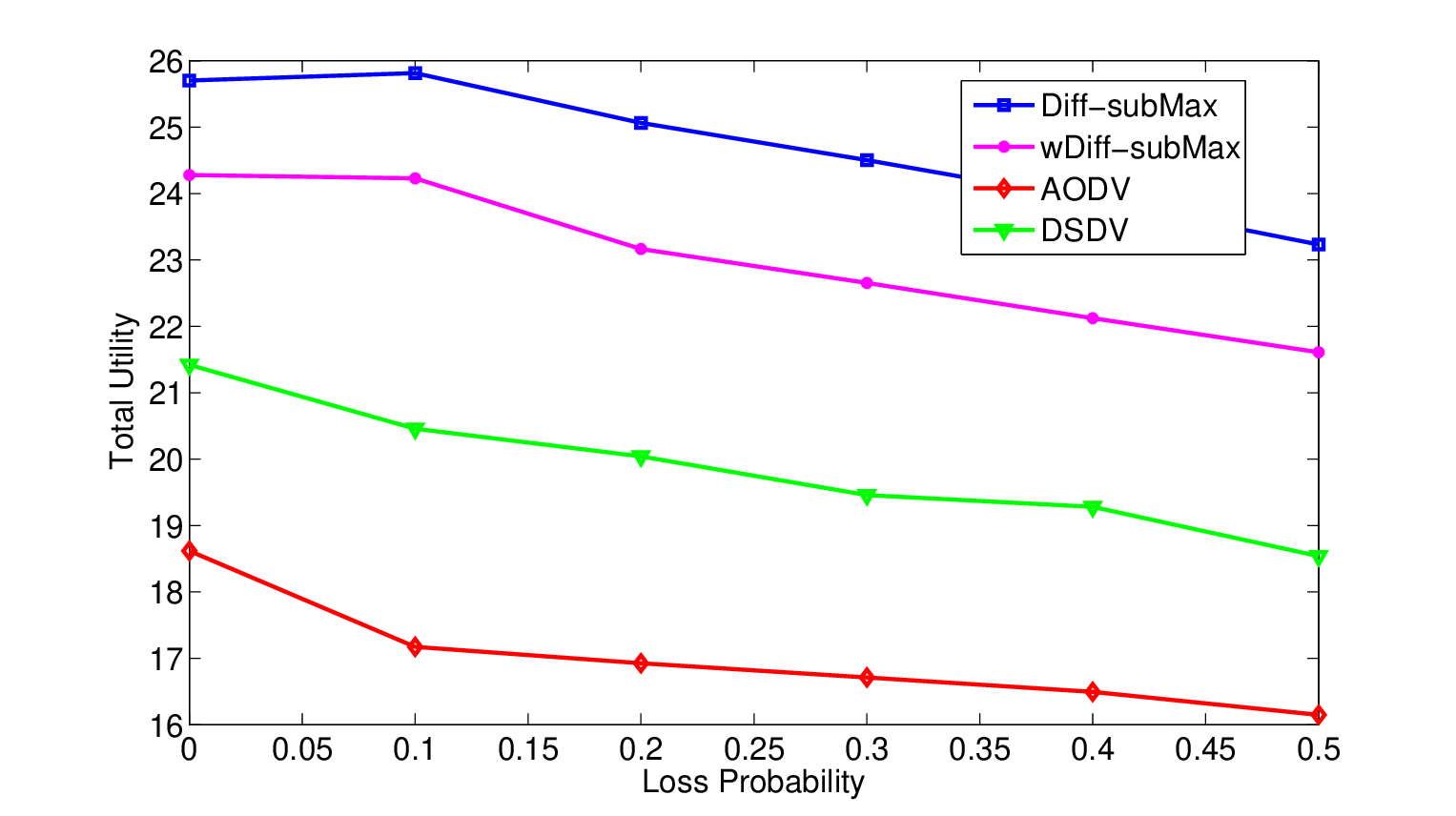}}} 
\subfigure[Delay.]{{\includegraphics[width=5.8cm]{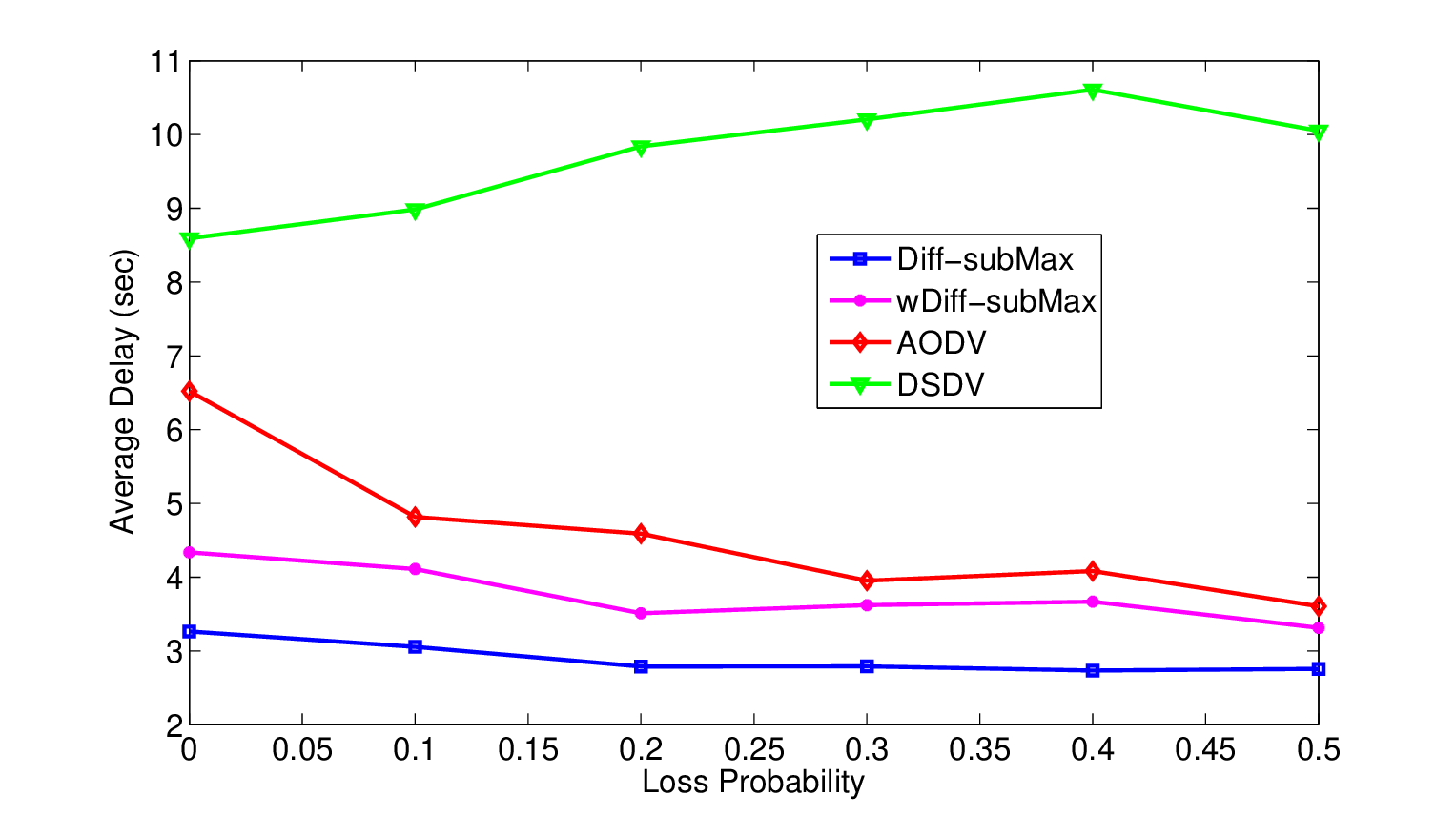}}} 
\end{center}
\begin{center}
\vspace{-10pt}
\caption{\label{fig:random_top_res2} Random Topology. (a) Total utility vs. average loss rate. (b) Average delay vs. average loss rate.}
\vspace{-10pt}
\end{center}
\end{figure}

Fig.~\ref{fig:random_top_res2} presents the simulation results for the random  topology. In this scenario, one third of the links, which are chosen randomly, are lossy with a rate between $0\%$ to $50\%$. Fig.~\ref{fig:random_top_res2}(a) shows the total utility vs. average loss rate results. As can be seen, our schemes significantly improve the total utility as compared to AODV and DSDV. The improvement in this scenario is higher as compared to the grid topology since there are more routing opportunities that can be exploited in this random topology scenario. 

Fig.~\ref{fig:random_top_res2}(b) shows the average delay vs. average loss rate results for the random topology. Diff-subMax and wDiff-subMax improve the delay performance as compared to AODV and DSDV. The improvement as compared to DSDV is especially significant. These results show that Diff-subMax and wDiff-subMax improve both utility and delay as compared to AODV and DSDV.

\subsection{Flow in the Middle Problem} 
In this section, we demonstrate the benefit of our modular algorithm design to address a specific problem called the ``flow in the middle problem'' \cite{DiffQ}.

\begin{figure}[t!]
\centering
 \scalebox{.7}{\includegraphics{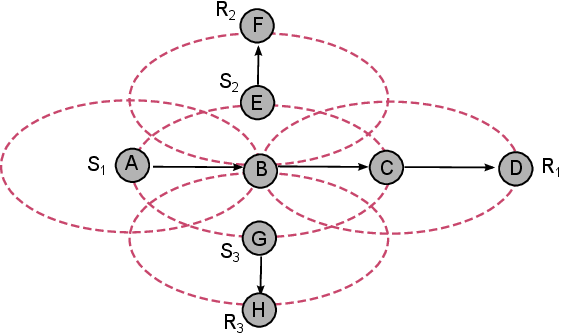}}
\caption{A topology that we consider for the flow in the middle problem. There are three flows; Flow $1$ from $S_1$ to $R_1$, Flow $2$ from $S_2$ to $R_2$, and Flow $3$ from $S_3$ to $R_3$. Each dashed ellipse shows a transmission and interference range of the node that is located in the center of the ellipse. In this scenario, Flow 1 suffers from the flow in the middle problem.}
\vspace{-15pt}
\label{fig:flow_in_the_middle_topology}
\end{figure}

Let us consider the topology shown in Fig.~\ref{fig:flow_in_the_middle_topology}, where there are three flows; Flow $1$ from $S_1$ to $R_1$, Flow $2$ from $S_2$ to $R_2$, and Flow $3$ from $S_3$ to $R_3$. In this scenario, Flow 1 suffers from the flow in the middle problem when 802.11 MAC is employed. In particular, node $B$ is subject to more interference as compared to node $E$ and $G$ as it shares the medium with four other nodes. On the other hand, nodes $E$ and $G$ share the wireless medium with only two nodes. Since, 802.11 MAC tends to give equal opportunity to all nodes in the wireless medium, node $B$ will transmit less, which will reduce the rate of Flow 1. Thus, the rates of Flow 2 and Flow 3 will be higher, while Flow 1 has lower rate. Note that, the flow in the middle problem arises from the fair share nature of 802.11 MAC. Since our Diff-subMax and wDiff-subMax algorithms also employ 802.11 MAC, the flow in the middle problem may also be seen in our algorithms.

However, thanks to our modular design, Diff-subMax could be easily updated to address the flow in the middle problem, which is not possible with AODV and DSDV. In particular, if the contention window size of 802.11 MAC is arranged depending on the weight of the chosen link (similar to \cite{DiffQ}, \cite{umut_stolyar}) in Diff-subMax, {\em i.e.}, if the contention window size is inversely proportional to $\omega_{l^{*}}$, where $l^{*}$ is the chosen link, then the middle in the flow problem can be alleviated.

\begin{figure*}[t!]
\centering
\subfigure[Diff-subMax with updated 802.11 MAC]{ \scalebox{.2}{\includegraphics{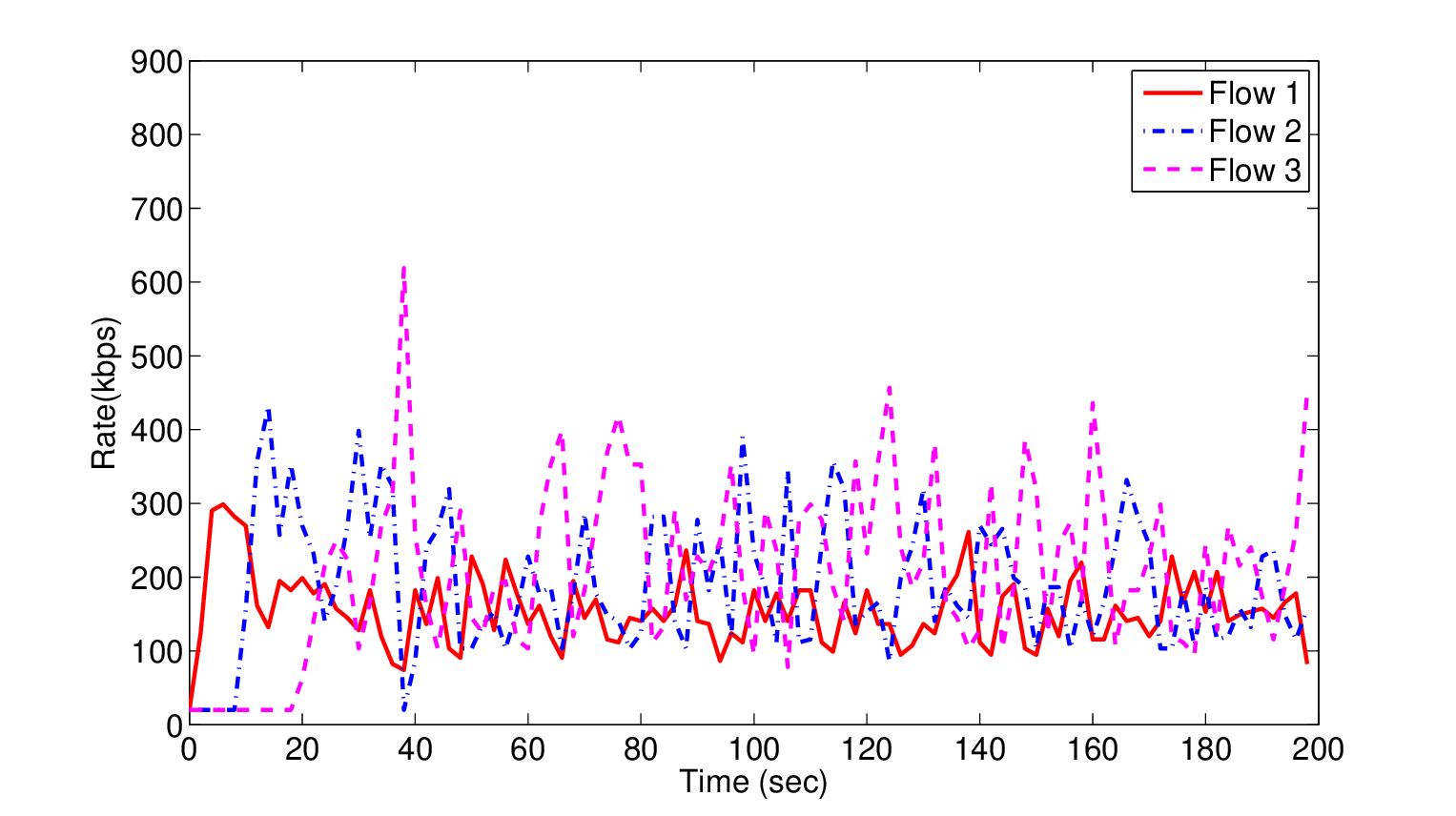}} }
\subfigure[AODV]{ \scalebox{.2}{\includegraphics{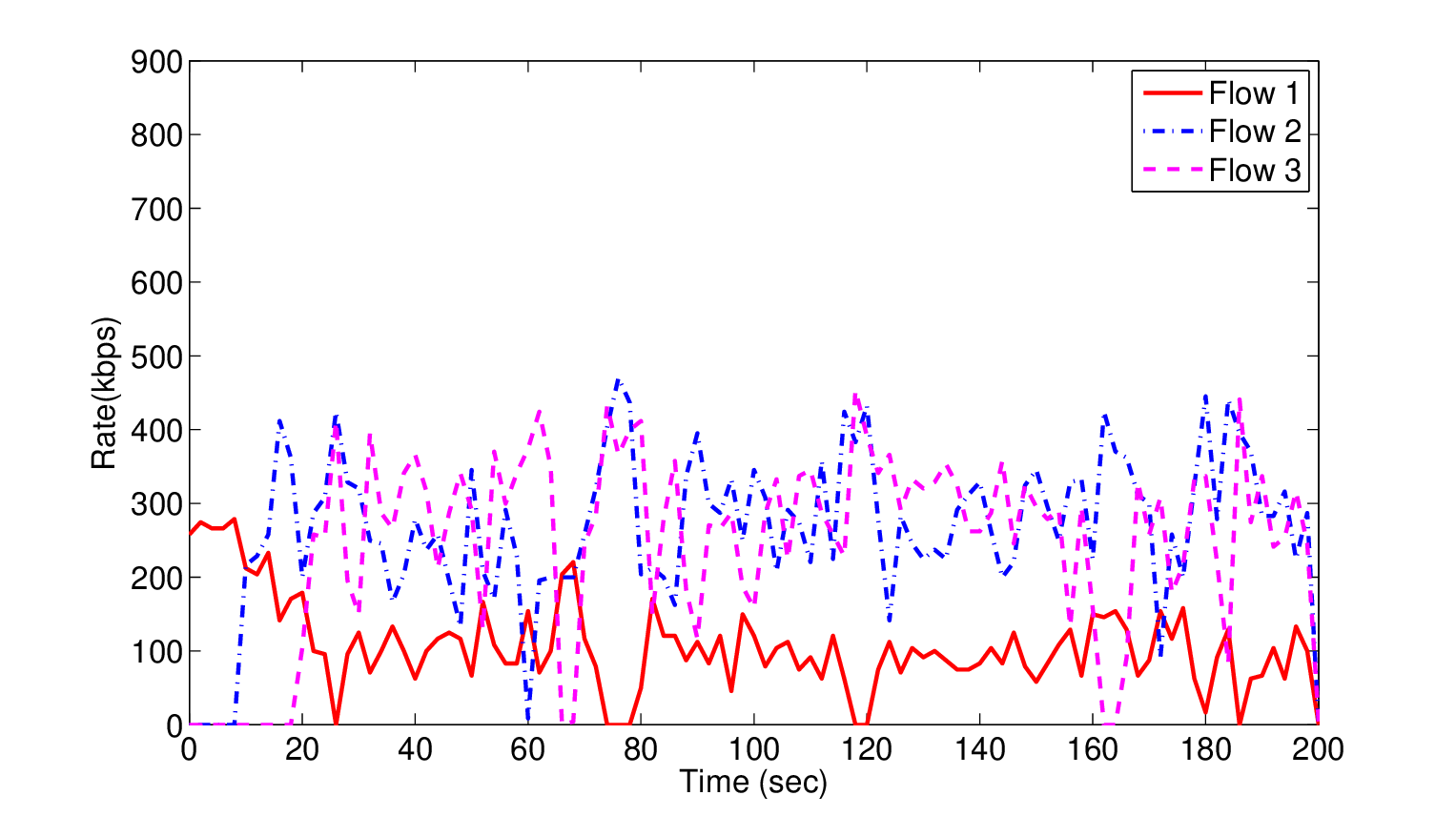}} }
\subfigure[DSDV]{ \scalebox{.2}{\includegraphics{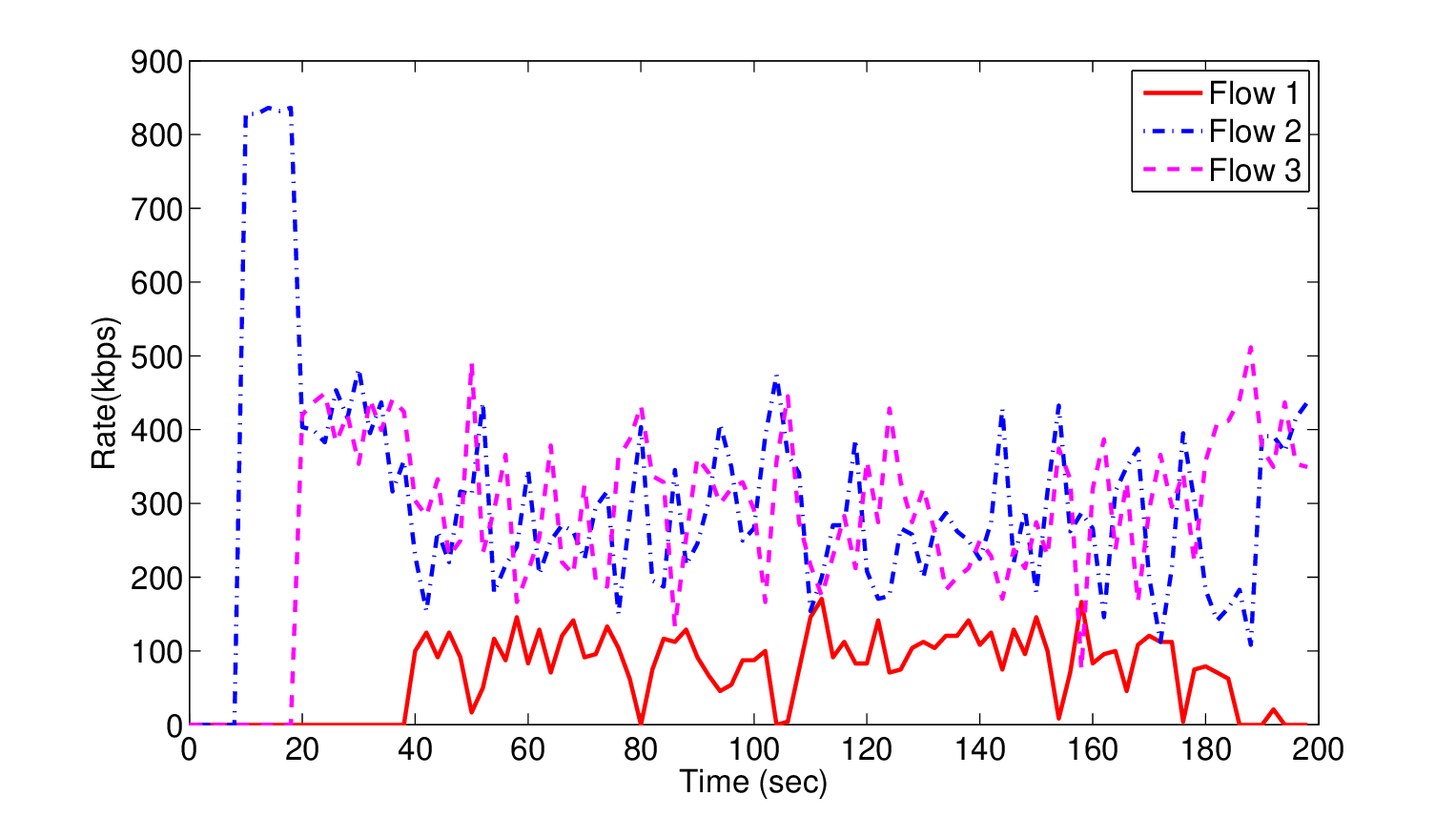}} }
\caption{Per-flow throughput vs time for the topology shown in Fig.~\ref{fig:flow_in_the_middle_topology}.  The channel capacity is $1Mbps$,  the buffer size at each node is set to $1000$ packets, packet sizes are set to $1000B$, and link loss probabilities are $0$. (a) Diff-subMax with updated 802.11 MAC. (b) AODV. (c) DSDV. }
\label{fig:flow_in_the_middle_results}
\end{figure*}

The ns-2 simulation results in Fig.~\ref{fig:flow_in_the_middle_results}(a) shows that Diff-subMax with updated 802.11 MAC addresses the flow in the middle problem for the topology shown in Fig.~\ref{fig:flow_in_the_middle_topology}. On the other hand, Flow $1$ still suffers in AODV and DSDV as shown in Fig.~\ref{fig:flow_in_the_middle_results}(b) and  Fig.~\ref{fig:flow_in_the_middle_results}(c). This shows the effectiveness of our modular design to address problems arising from different layers.

\section{\label{sec:related}Related Work}
{\em Backpressure and Follow-up Work:} This paper builds on backpressure, a routing and scheduling framework for communication networks \cite{tass_eph1}, \cite{tass_eph2}, which has generated a lot of interest in the research community \cite{neely_book}; especially for wireless ad-hoc networks \cite{tass3}, \cite{kahale}, \cite{andrews}, \cite{neely_mod_pow}, \cite{stolyar_greedy}, \cite{liu_stolyar}. Furthermore, it has been shown that backpressure can be combined with flow control to provide utility-optimal operation guarantee \cite{neely_mod}, \cite{stolyar_greedy}. This paper follows the main idea of backpressure, and revisits it considering the practical challenges that are imposed by current networks.

{\em Backpressure Implementation:} The strengths of backpressure have recently increased the interest in the practical implementation of backpressure over wireless networks. Multi-path TCP scheme is implemented over wireless mesh networks in \cite{horizon} for routing and scheduling packets using a backpressure based heuristic.
At the link layer, \cite{DiffQ}, \cite{umut_stolyar}, \cite{sridharan1}, \cite{sridharan2} propose, analyze, and evaluate link layer backpressure-based implementations with queue prioritization and congestion window size adjustment. Backpressure is implemented over sensor networks \cite{routing_wtht_routes} and wireless multi-hop networks \cite{xpress}, which are also the closest implementations to ours. The main difference in our work are that; (i) we consider separation of routing and scheduling to make practical implementation easier, (ii) we design and analyze a new scheme; Diff-Max, (iii) we simulate and implement Diff-Max over ns-2. 

{\em Backpressure and Queues.} According to backpressure, each node constructs per-flow queues. There is some work in the literature to stretch this necessity. For example, \cite{pkt_by_pkt_adap_rout}, \cite{locbui} propose using real per-link and virtual per-flow queues. Such a method reduces the number of queues required in each node, and reduces the delay, but it still makes routing and scheduling decisions jointly and does not separate routing from scheduling. 
Therefore, this approach requires strong synchronization between the network and link layers, which is difficult to implement in practice as explained in Section~\ref{sec:intro}.

\section{\label{sec:conclusion}Conclusion}
In this paper, we proposed Diff-Max, a framework that separates routing and scheduling in backpressure-based wireless networks. The separation of routing and scheduling makes practical implementation easier by minimizing cross-layer operations and it leads to modularity. Our design is grounded in the network utility maximization (NUM) formulation of the problem and its solution. Based on the structure of Diff-Max, two practical schemes, Diff-subMax and wDiff-subMax are developed. Simulations in ns-2 demonstrate significant improvement in terms of throughput, utility, and packet delay as compared to AODV and DSDV.

\bibliographystyle{IEEEtran}

\section*{\label{sec:appendixA}Appendix A: Analysis of Deterministic Solutions}
\subsection{Diff-Max} In this section, we analyze the deterministic solution of Diff-Max. We first, explain the evolution of Lagrange multipliers, and then discuss the convergence of the solution to the optimal point. 

{\em Diff-Max - Lagrange Multipliers:} The Lagrange multipliers; $u_{i}^{s}$ and $v_{i,j}$ are calculated using gradient descent:
\begin{align} \label{opt:eq1_parameterUpdate}
u_{i}^{s}(t+1) = & \{ u_{i}^{s}(t) - \alpha_t [ \sum_{j \in \Nset} f_{i,j}^{s}(t) - \sum_{j \in \Nset} h_{j,i}^{s}(t) \nonumber \\
& - x_{s}(t)1_{i=o(s)} ] \}^{+} \nonumber \\
v_{i,j}(t+1) = & \{v_{i,j}(t) + \beta_t [ \sum_{s \in \Sset} f_{i,j}^{s}(t) - h_{i,j}(t) ] \}^{+}
\end{align} where $t$ is the iteration number, $\alpha_t$ and $\beta_t$ are the step sizes of the gradient descent algorithm, and the $~^+$ operator makes the Lagrange multipliers positive.

{\em Diff-Max - Convergence to the Optimal Point:} The convergence of the distributed solutions of Diff-Max, \ie Eqs.~(\ref{opt:eq1_rateControl}), (\ref{opt:eq1_routing2}), (\ref{opt:eq1_scheduling1}), (\ref{opt:eq1_parameterUpdate}) follows directly from the convergence of convex optimization problems through gradient descent \cite{lin_schroff_paper}, \cite{bert_tsit_book}. In particular, if $\lim_{t\rightarrow\infty} \alpha_t = 0, \sum_{t=0}^{\infty}\alpha_t=\infty$ and $\lim_{t\rightarrow\infty} \beta_t = 0, \sum_{t=0}^{\infty}\beta_t=\infty$, then the solution converges, \ie $\lim_{t\rightarrow\infty} \| x(t) - x^{*}\| = 0$, where $x^{*}$ is the utility optimal operating point of the convex optimization problem in (\ref{opt:eq1}).

\subsection{Diff-subMax}
In this section, we analyze Diff-subMax and its convergence properties.

{\em Diff-subMax - Problem Formulation:} Since Diff-subMax uses 802.11's CSMA/CA mechanism for inter-node scheduling, it is a solution to the following problem. 
\begin{align} \label{opt:diffsubMax}
\max_{\boldsymbol x, \boldsymbol f, \boldsymbol h} & \sum_{s \in \Sset} g_{s}(x_{s}) \nonumber \\
\mbox{s.t. } & \sum_{j \in \Nset} f_{i,j}^{s} - \sum_{j \in \Nset} h_{j,i}^{s} = x_{s}1_{[i=o(s)]}, \forall i \in \Nset, s \in \Sset \nonumber \\
& \sum_{s \in \Sset} f_{i,j}^{s} \leq h_{i,j}, \forall (i,j) \in \Lset \nonumber \\
& f_{i,j}^{s} = h_{i,j}^{s}, \forall s \in \Sset, (i,j) \in \Lset \nonumber \\
& h_{i,j} \leq (1-\bar{p}_{i,j})\bar{R}_{i,j} \tau_{i,j}, \forall i \in \Nset, j \in \Nset \nonumber \\
& \sum_{j \in \Nset} \tau_{i,j} \leq \gamma_{i}, \forall i \in \Nset
\end{align} where $\bar{p}_{i,j}$ and $\bar{R}_{i,j}$ are the loss probability and link rate over link $i-j$, $\tau_{i,j}$ is the percentage of time that link $i-j$ is used for transmission, and $\gamma_{i}$ is the percentage of time node $i$ is active for transmission. The percentage of time that a node is active, \ie $\gamma_{i}$ is determined by CSMA/CA, and it is constant in our problem. Note that the only difference of (\ref{opt:diffsubMax}) as compared to (\ref{opt:eq1}) are the last two constraints. In particular, since CSMA/CA is employed for inter-node scheduling, it gives opportunity to each node for transmission. The percentage of these transmission opportunities is constant in our problem, because CSMA/CA makes these decisions independent from our routing and inter-node scheduling decisions. Then, after node $i$ is given opportunity for transmission by CSMA/CA, we determine the best link $i-j$ to activate. Thus, the sum of the percentages of per-link activations; \ie $\sum_{j \in \Nset} \tau_{i,j}$  should be less than the percentage of constant node activation; \ie $\gamma_{i}$, as shown in the last constraint of (\ref{opt:diffsubMax}). Also, the percentages of link activations can be translated into the link transmission rates as shown in the fourth constraint of  (\ref{opt:diffsubMax}).

{\em Diff-subMax - Decomposed Solution:} If we solve the NUM problem in (\ref{opt:diffsubMax}), we get the same flow control, and routing problems as in Diff-Max, as we also explained in Section~\ref{sec:algs_diffsubMax}. In particular, the flow control part solves
\begin{equation} \label{opt:eq1_rateControl_diffsubMax}
\textstyle x_s = ({g'_{s}})^{-1} \left( u_{o(s)}^{s} \right),
\end{equation} as in (\ref{opt:eq1_rateControl}), and the routing part solves 
\begin{align} \label{opt:eq1_routing2_diffsubMax}
\max_{\boldsymbol f} & \sum_{(i,j) \in \Lset} \sum_{s \in \Sset} f_{i,j}^{s} (u_{i}^{s} - u_{j}^{s} - v_{i,j})
\end{align} as in (\ref{opt:eq1_routing2}). On the other hand, the scheduling part changes as it solves; 
\begin{align} \label{opt:eq1_scheduling1_diffsubMax1}
\max_{\boldsymbol h} & \sum_{(i,j) \in \Lset} v_{i,j} h_{i,j} \nonumber \\
\mbox{s.t. } & h_{i,j} \leq (1-\bar{p}_{i,j})\bar{R}_{i,j} \tau_{i,j}, \forall i \in \Nset, j \in \Nset \nonumber \\
& \sum_{j \in \Nset} \tau_{i,j} \leq \gamma_{i}, \forall i \in \Nset
\end{align} This problem is expressed as
\begin{align} \label{opt:eq1_scheduling1_diffsubMax2}
\max_{\boldsymbol h} & \sum_{(i,j) \in \Lset} v_{i,j} (1-\bar{p}_{i,j})\bar{R}_{i,j} \tau_{i,j} \nonumber \\
\mbox{s.t. } & \sum_{j \in \Nset} \tau_{i,j} \leq \gamma_{i}, \forall i \in \Nset
\end{align} This problem is equivalent to determining the link $l^{*}$ according to $l^* = \argmax_{[j \in \Nset_{i}]} \omega_{i,j}$, where $\omega_{i,j}=v_{i,j}(1-\bar{p}_{i,j})\bar{R}_{i,j}$. This solution is the deterministic version of what we proposed to implement in Section~\ref{sec:algs_diffsubMax}. Next, we discuss the Lagrange multipliers of Diff-subMax.

{\em Diff-subMax - Lagrange Multipliers:} The Lagrange multipliers; $u_{i}^{s}$ and $v_{i,j}$ are calculated using gradient descent:
\begin{align} \label{opt:eq1_parameterUpdate_diffsubMax}
u_{i}^{s}(t+1) = & \{ u_{i}^{s}(t) - \alpha_t [ \sum_{j \in \Nset} f_{i,j}^{s}(t) - \sum_{j \in \Nset} (1-\bar{p}_{j,i})\bar{R}_{j,i} \tau_{j,i}^{s}(t) \nonumber \\
& - x_{s}(t)1_{i=o(s)} ] \}^{+} \nonumber \\
v_{i,j}(t+1) = & \{v_{i,j}(t) + \beta_t [ \sum_{s \in \Sset} f_{i,j}^{s}(t) - (1-\bar{p}_{i,j})\bar{R}_{i,j} \tau_{i,j}(t) ] \}^{+}
\end{align} where $t$ is the iteration number, $\alpha_t$ and $\beta_t$ are the step sizes of the gradient descent algorithm, the $~^+$ operator makes the Lagrange multipliers positive, and $\tau_{j,i}^{s}$ is the percentage of time that link $j-i$ is used for transmitting packets from flow $s$.

{\em Diff-subMax - Convergence to the Optimal Point:} The convergence of the solution set, Eqs.~(\ref{opt:eq1_rateControl_diffsubMax}), (\ref{opt:eq1_routing2_diffsubMax}), (\ref{opt:eq1_scheduling1_diffsubMax2}), (\ref{opt:eq1_parameterUpdate_diffsubMax}) follows directly from the convergence of convex optimization problems through gradient descent \cite{lin_schroff_paper}, \cite{bert_tsit_book}. In particular, if $\lim_{t\rightarrow\infty} \alpha_t = 0, \sum_{t=0}^{\infty}\alpha_t=\infty$ and $\lim_{t\rightarrow\infty} \beta_t = 0, \sum_{t=0}^{\infty}\beta_t=\infty$, then the solution converges, \ie $\lim_{t\rightarrow\infty} \| x(t) - x^{*}\| = 0$, where $x^{*}$ is the optimal solution to (\ref{opt:diffsubMax}). Note that the optimal solution of Diff-subMax could be smaller than the optimal solution of Diff-Max, because Diff-Max also optimizes inter-node scheduling, while Diff-subMax uses CSMA/CA for inter-node scheduling. However, Diff-subMax still optimizes flow control, routing, and intra-node scheduling, and its deterministic version converges to the utility optimal operating point.

\end{document}